\documentclass[twocolumn]{aastex63}
\usepackage{lineno}
\received{\today}
\submitjournal{ApJ}

\shorttitle{SN-driven cusp-core transformation}
\shortauthors{Burger and Zavala}
\graphicspath{{./}{figures/}}

\begin{document}

\title{SN-driven mechanism of cusp-core transformation: an appraisal}

\correspondingauthor{Jan Burger}
\email{jdb5@hi.is}

\author[0000-0001-8293-3709]{Jan D. Burger}
\affiliation{Centre for Astrophysics and Cosmology, Science Institute, University of Iceland, Dunhagi 5, 107 Reykjavik, Iceland}

\author[0000-0003-4442-908X]{Jesús Zavala}
\affiliation{Centre for Astrophysics and Cosmology, Science Institute, University of Iceland, Dunhagi 5, 107 Reykjavik, Iceland}










\begin{abstract}
We present and test 
an effective model for $N$-body simulations that aims at mimicking
the impact of supernova (SN) feedback on the dark matter (DM) distribution of isolated halos hosting dwarf galaxies. 
Although the model is physically decoupled from the cosmological history of both the DM halo and the dwarf galaxy, it allows us to 
study the impact of different macroscopic parameters 
such as galaxy concentration, feedback energy and energy injection time in the process of SN-driven core formation in a physically clear way.
Using our effective model in a suite of $N$-body simulations of an isolated halo with different 
SN feedback parameters, we find that whether or not a DM core forms 
is determined by the total amount of SN feedback energy that is transferred to the DM particles. At a fixed injected energy, the amount of transferred energy is bigger -- and the size of the DM core is larger -- the faster the energy injection occurs and the more compact the dwarf galaxy is. Analyzing the orbital evolution of kinematic tracers, we demonstrate that a core forms through SN feedback only if the energy injection is impulsive relative to the dynamical timescale of particles in the inner halo. However, there is no fundamental link between the total amount of injected energy and the injection rate.
Consequently, the presence of signatures of impulsive changes of the gravitational potential is not a sufficient condition for dwarf-size halos to have cored density profiles.   
\end{abstract}



\section{Introduction} \label{sec:intro}

The overall successful cold dark matter (CDM) paradigm of structure formation faces potentially severe challenges on small, sub-galactic scales (see \citealt{Bullock2017} for a review), particularly in its comparison with the properties of the population of satellites in the Milky Way (MW). For instance, inferred circular velocities of the 
MW satellites are inconsistent with the properties of the subhalos found in MW-like halos identified in dark matter only (DMO) CDM cosmological simulations; a mismatch that is known as the too-big-to fail (TBTF) problem (\citealt{Boylan-Kolchin2011,BK2012}). 
An updated (extended) version of the TBTF problem 
highlights that MW satellites reside within DM subhalos with a distribution of circular velocity profiles that is too diverse to be easily reconciled with the narrow distribution seen in (DMO) CDM simulations \citep{2019PhRvD.100f3007Z}. 

A second, possibly related, problem of the DMO CDM model is that it predicts halos to have central density cusps, which is potentially in disagreement with observations of some dwarf galaxies.
This so-called cusp-core problem has been put forward e.g. through observations of HI-rotation curves of field dwarfs (e.g. \citealt{Moore1994}, \citealt{deBlok2008}, \citealt{Kuzio2008}, \citealt{2019MNRAS.484.1401R}) and through a kinematic analysis of two stellar populations of different age in the dwarf spheroidal galaxies Fornax and Sculptor \citep{Walker2011}. However, there is some debate about whether these observations truly imply cored and isothermal DM density profiles. For instance, \citet{2019MNRAS.482..821O} argue that non-circular motion can lead to a large diversity in rotation curves (and hence to a misleading inference of the presence of a cusp or a core), depending on the observer's line of sight (see also \citealt{2020MNRAS.495...58S}), while \citet{2018MNRAS.474.1398G} argue that the measurement by \citet{Walker2011} could potentially be flawed due to a violation of spherical symmetry in Fornax and Sculptor. While the evidence of the presence of the ubiquity of cored density profiles remains controversial, 
several mechanisms of cusp-core transformation
have been investigated. New DM physics can, for instance, potentially solve both the TBTF and the cusp-core problem simultaneously (see \citealt{2019Galax...7...81Z} and references therein). As the kinematics of dwarf galaxies are largely determined by their DM content, DM candidates which deviate from the cold and collision-less CDM hypothesis -- and offer solutions to CDM's small scale problems -- should be considered a feasible alternative to CDM, provided they manage to evade constraints derived from observations on larger cosmological scales. 
One such candidate is self-interacting dark matter (SIDM, \citealt{Spergel2000}, \citealt{Yoshida2000}, \citealt{Vogelsberger2012}, \citealt{Dave2001}, \citealt{Colin2002}, \citealt{Rocha2013}). In SIDM halos, elastic scattering between individual DM particles causes a redistribution of energy from the outer parts of the halo into its center, creating an isothermal core in a fully adiabatic way. SIDM is a feasible DM candidate if the self-interaction cross section evades current astrophysical constraints (e.g. from elliptical galaxies \citealt{Peter2013}, clusters \citealt{Robertson2017,Robertson2018}, and dwarf galaxies 
\citealt{Read2018}). 

Given that the potential CDM challenges at small scales are only strongly supported from comparisons between observations and results of DMO simulations, the idea that an adequate modelling of baryonic physics may provide a solution is appealing. 
In particular, supernova feedback has been invoked to solve both the TBTF and cusp-core problems.
\citet{Navarro1996} showed that the sudden removal of an 
external disk potential can cause initially cuspy halos to form cores in DMO simulations. 
\citet{Gnedin2002} repeated the experiment
concluding that a single mass removal equivalent to one entire galaxy
is not sufficient to trigger the formation of isothermal cores despite significantly reducing the inner halo density. 
A decade later, \citet{Pontzen:2011ty} developed a model to explain core formation through SN feedback as observed in modern hydrodynamical simulations. They show that repeated SN-driven outflows of gas can cause halos to form a core, provided that 
the energy injection occurs much faster than the typical dynamical timescale in the inner part of the halo. From observations, there is some evidence that star formation histories in dwarf galaxies at the high mass are indeed ``bursty'' (and the subsequent supernovae feedback cycles impulsive), i.e., they happen on timescales that 
are comparable to the dynamical time of the galaxy 
\citep{Kauffmann2014}, but 
observations have yet to reach the time resolution needed
to resolve the starbust cycle on timescales smaller than the dynamical timescales of the (low-mass) MW dwarf satellite population 
\citep{Weisz2014}. Based on \citet{Pontzen:2011ty}'s idea of core formation through periodic SN-driven mass removal, \citet{GarrisonKimmel:2013aq} slightly altered the model used by \citet{Navarro1996} and showed that SN feedback alone cannot solve the classical TBTF problem (not taking into account ultra faint dwarfs). A general discussion of the coupling between the cusp-core problem and the TBTF problem was presented in \citet{Penarrubia:2012bb}, where, most notably, the authors calculate the energy required to form a core as a function of halo mass. 

From hydrodynamical simulations, there is growing consensus that cores can form as a result of episodic impulsive SN feedback, and that whether or not they do form depends on the ratio of stellar mass to halo mass in a given halo (\citealt{DiCintio2014}, \citealt{Tollet2016}, \citealt{Chan2015}, \citealt{2017MNRAS.471.3547F}, \citealt{2020MNRAS.497.2393L}). In particular, the universal finding is that there is a limited range of stellar to halo mass ratios for which SN feedback is efficient at forming cores, corresponding for the most part to the mass range of bright/massive dwarfs (see, e.g. Figure 2 of \citealt{2020MNRAS.497.2393L}). 
Performing high resolution hydrodynamic simulations of isolated halos, \citet{Read2016} find that SN feedback can cause core formation even in ultra faint dwarfs if star formation is sustained for long enough (see also \citealt{Amorisco2014} for the relevant impact of star formation histories on the plausibility of forming cores in MW satellites). The origin of this disagreement is not entirely clear, but a possible reason may be that the baryon fraction and/or star formation histories assumed in the controlled/isolated simulations of \citet{Read2016} cannot be realized in a full (CDM) cosmological setting, or vice versa, the latter assumptions are present in nature but cannot yet be obtained/modelled in full (CDM) cosmological simulations.
To date, the question of whether or not SN feedback can cause core formation in ultra faint dwarfs remains a subject of debate (see e.g. \citealt{2020arXiv201109482G}, \citealt{2019MNRAS.490.4447W}, \citealt{2021arXiv210102688O}). 

Two commonly identified conditions for core formation in the hydrodynamical simulations mentioned above are i) that the energy condition of \citet{Penarrubia:2012bb} for cusp-core transformation must be fulfilled and ii) that the dwarf galaxies have a sufficiently bursty star formation history, leading to gaseous outflows, and thus fluctuations of the gravitational potential, on rather short timescales. However, \citet{2019MNRAS.486.4790B} bring attention to the fact that no cores are formed in dwarfs within the AURIGA and APOSTLE simulations, despite their simulated dwarf galaxies' bursty star formation history. The solution, as presented in \citet{2019MNRAS.488.2387B}, is that whether or not cores form through SN feedback also depends on the numerical value of the star formation threshold adopted in the simulations, a fact that had already been discussed in \citet{Pontzen:2011ty}. \citet{2019MNRAS.488.2387B} state that larger star formation thresholds lead to higher gas densities in the inner halo, eventually causing the gas to dominate the gravitational potential in the halo's center. As SN feedback couples to DM solely through the gravitational interaction between DM and baryons, the impact of SN feedback is maximized if the contribution of gas to the gravitational potential is significant, as SN feedback would then cause stronger fluctuations in the potential.  While \citet{2019MNRAS.488.2387B} find that there is a range of star formation threshold values which corresponds to a ``sweet spot'' for core formation, \citet{2020MNRAS.499.2648D} argue that the final DM profiles of dwarf-sized halos converge towards having an isothermal core for ever larger numerical star formation thresholds, provided that the right softening length is chosen in the simulation.  

Therefore, it appears that whether or not cores are formed in hydrodynamical simulations 
of dwarf galaxies depends on three different macroscopic conditions:
\begin{itemize}
\item Is the injected feedback energy sufficient to transform a cuspy profile into a cored one?
\item Are the baryons sufficiently concentrated in the halo center prior to the first starburst?
\item Is the galaxy's star formation history bursty enough to cause energy injection on timescales smaller than the dynamical time in the halo center?
\end{itemize}
The answer to these questions depends on a complex regulation of different properties in
hydrodynamical simulations, which are ultimately limited by subgrid (unresolved) physics, most notably,
the (effective) star formation threshold and the implementation of SN feedback itself (its local energy/momentum deposition in the surrounding gas elements/particles). 
In this article, we follow the philosophy of 
\citet{GarrisonKimmel:2013aq} and test the impact of changing these three properties directly in a controlled way by 
introducing an effective model for SN feedback, consisting of an external potential mimicking the gravitational pull of a galaxy, as well as a scheme to periodically inject energy into the halo. In our scheme, the spatial distribution of ``supernovae'' is determined by the mass distribution corresponding to the external potential and the time over which energy is injected into the halo is free parameter. This simple model allows us to separate the effects of the three key properties connected to core formation, enabling a more transparent physical interpretation than in full hydrodynamical simulations.

In a suite of DMO simulations of an isolated DM halo with identical initial conditions, we look at the effects of including three different (external) Plummer galaxies of equal mass but with different half-light radii, as well as three different exponential disk galaxies, also of equal mass but with different half-light radii. We furthermore vary the total injected energy, and the time over which the energy is injected. To analyze whether our SN feedback scheme is adiabatic or impulsive compared to the dynamical timescale in the center of the halo, we monitor the evolution of the phase space density of an orbital family; a method that was first introduced in \citet{2019MNRAS.485.1008B}. 

This article is structured as follows. We present our effective model for supernova feedback in Section \ref{sec:method}. The simulations we ran to test our model are introduced and discussed in Section \ref{sec:simulations}. The main results are presented in Section \ref{sec:results} while their implications are discussed in Section \ref{sec:discussion}, along with a discussion of some of the approximations made in our model. Finally, we summarize in Section \ref{sec:conclusion}.

\section{Method}\label{sec:method}
Our effective model of supernova feedback is built upon the one introduced in \citet{2019MNRAS.485.1008B}. Therein, supernova feedback 
was modelled as a periodic addition and subsequent sudden removal of an external potential located at the halo's center. 
We modelled the external potential as a \citet{Hernquist:1990be} sphere and thus, in a coordinate system with origin at the halo's center, including the extra potential amounts to an additional external acceleration for each particle in the simulation:
\begin{equation}
    \mathbf{a}_{\rm ext} = -\frac{GM(t)}{(r+r_s)^2}\frac{\mathbf{r}}{r},
\end{equation}
where $M(t)$ is a function defining the time-dependent mass corresponding to the external potential and $r_s$ is the scale length of the Hernquist sphere. While this model serves the purpose of approximating a very sudden central -- star-burst like -- injection of energy into the halo, it is clear that it serves as a rather coarse approximation to the true effects of SN feedback. Nonetheless, it is a time-efficient method to investigate kinematic signatures of tracer particles in a halo which undergoes impulsive core formation - without the need to perform full hydrodynamical simulations. 
In this article, we aim to expand upon the \citet{2019MNRAS.485.1008B} method with two distinct goals. Firstly, we would like to model 
the dependence of SN feedback on a set of three relevant parameters that can be connected to both observations and hydrodynamical simulations: 
galaxy size, total feedback energy, 
and the time over which the SN energy deposition occurs (hereafter to be called injection time). Secondly, we attempt to couple the SN feedback energy release to the energy budget that is actually available in a galaxy of the modelled size. 

In this Section we describe our improved method by first outlining how the external potential is placed within the center of the halo. Then we present our implementation of both a spherically symmetric external Plummer potential and the external potential generated by an axisymmetric flat exponential disk. Finally, we discuss our implementation of SN feedback-like energy input and how we relate it to the total stellar mass and the size of observed dwarf galaxies.

\subsection{Determination of the halo's center of potential}\label{sec:cop}

To place our external, galaxy-mimicking, potential in such a way that its position is consistent with the halo's self-gravitating potential, we need to calculate the halo's center of potential at each time-step. We do this in an iterative manner using a shrinking spheres method. As we are concerned with a single isolated halo in this work, our first step takes into account all DM particles in the simulation\footnote{We note that the method can easily be extended to halos within a larger simulation that have been identified by, for instance, a friends-of-friends algorithm.}. In the first step, the location of the center of potential is estimated as 
\begin{equation}
    \mathbf{R} = \frac{\displaystyle\sum_{i=1}^{N}\mathbf{r}_i \Phi(\mathbf{r}_i)}{\displaystyle\sum_{i=1}^{N}\Phi(\mathbf{r}_i)},\label{eq:basic_cop}
\end{equation}
where $N$ is the number of DM particles in the halo, $\mathbf{r}_i$ is each particle's position vector, $\Phi(\mathbf{r}_i)$ is the potential at the particle's position and it is understood that the potential is defined such that 
\begin{equation}
    \Phi(\mathbf{r}) = -\sum_{i=1}^{N}\frac{Gm_i}{|\mathbf{r}_i-\mathbf{r}|},
\end{equation}
where $m_i$ is the mass of the {\it i}'th DM particle. After calculating an initial estimate of $\mathbf{R}$, we repeat the calculation in Equation (\ref{eq:basic_cop}), limiting the sum to particles for which $|\mathbf{r}_i-\mathbf{R}_{\rm e}|< r_{\rm target}$, where $\mathbf{R}_{\rm e}$ is the center of potential estimated in the previous step, and $r_{\rm target}$ is a target radius which we decrease each iteration. In this paper we use three values $r_{\rm target} = 50\,h^{-1}{\rm kpc},\,5\,h^{-1}{\rm kpc~and}\,0.5\,h^{-1}{\rm kpc}$. The result of the last iteration is then taken to be the halo's center of potential. 
During the last iteration, we also calculate the velocity of the halo's center of potential in analogy to Equation \ref{eq:basic_cop}, restricting the sum as outlined above and replacing the particles' position vectors with their velocities. We note that the implementation of our effective model of SN feedback does not require knowledge of the velocity of the halo's center of potential. However, we do use the velocity to place the halo's center ``at rest'' when determining the phase space distribution of tracer particles (see Section \ref{sec:results}).

\subsection{External potentials}\label{sec:extpot}
In a more realistic setting, 
the total amount of energy injected and coupled to the DM due to SN feedback 
depends on the amount and distribution of stellar mass within the DM halo. 
This stellar mass, however, can also cause an adiabatic contraction of the DM halo. When investigating core formation due to SN feedback, it is important to take this effect into account, as it can counteract, at least in part, the cusp-core transformation triggered by an impulsive energy injection. We model the net effect of a baryonic component by including an external potential into the simulation, centered at the halo's center of potential. We examine two cases which are of importance in dwarf-sized halos, a spherically symmetric Plummer potential (to mimic a bulge/spheroid) and the potential of an axisymmetric, flat exponential disk. 

\subsubsection{An external Plummer sphere}\label{sec:plpot}
The gravitational effect of a spherically symmetric Plummer profile 
can be approximated by adding an acceleration to each particle: 
\begin{equation}\label{eq:extplummer}
\mathbf{a}_{i,{\rm ext}} = -\frac{GM_{\rm Pl}}{\left({a^2+(\mathbf{r}_i-\mathbf{R})^2}\right)^{3/2}}\left(\mathbf{r}_i - \mathbf{R}\right),
\end{equation}
where $M_{\rm Pl}$ is the total mass of the galaxy modelled by the external potential and $a$ is the scale length of the Plummer sphere. To connect our model to observations of dwarf galaxies, we note that the half-mass radius of the Plummer profile, which is $r_{1/2} \sim 1.3 a$, can be compared to observed half-light radii. 

\subsubsection{An external axisymmetric flat disk}\label{sec_diskpot}
Including an analytic potential to model a disk galaxy is a somewhat more complicated task. In fact, for a vertically extended disk, the calculation of the external acceleration generated by the disk cannot be solved analytically. For that reason, we here model the disk to be infinitely flat, in which case the Poisson equation can be solved up to an integral and the vertical and radial force components can be calculated through numerical differentiation of the potential generated by the disk. We further assume the disk to be homogeneous and thus, the force generated by the disk has no azimuthal component. The flat disk is then fully characterized by its surface density profile 
\begin{equation}
    \Sigma(R) = \frac{M_{\rm d}}{2\pi H^2}\exp\left(-\frac{R}{H}\right),\label{eq:disk_surface}
 \end{equation}
 with $M_{\rm d}$ being the total mass of the disk, $H$ its scale length and $R$ is the polar (cylindrical) radius. The total mass (volume) density is $\rho(r) = \Sigma(R)\delta_D(z)$, and thus we can solve Poisson's equation to obtain the potential: 
 \begin{equation}
     \Phi(R,z) = -\int_0^\infty dk\,\frac{GM_{\rm d}J_0(kR)\exp(-k|z|)}{\sqrt{1+(kH)^2}^3}.\label{eq:disk_pot}
 \end{equation}
In Equation \ref{eq:disk_pot}, $J_0(kR)$ is a Bessel function of the first kind. We evaluate Equation \ref{eq:disk_pot} at the start of our simulations on a grid of $(R,z)$ values after testing whether the integral is converged by systematically varying the upper integration limit. During the simulation, the potential at a given point in space can then be calculated through interpolation over the values calculated at the grid points. The force at a certain point in space is then easily obtained as the (directional) numerical derivative of this two-dimensional interpolated potential.

\begin{deluxetable*}{lccl}
\tablenum{1}
\tablecaption{Parameters of the effective model for supernova feedback (SNF; see Section \ref{sec:method}). \label{tab:model_params}}
\tablewidth{0pt}
\tablehead{
\colhead{Galaxy (external potential)} & \colhead{Parameter} & \colhead{Units} & \colhead{Description}
}
\startdata
  Plummer/Disk & $\epsilon$ & - &  Nominal SN energy coupling efficiency \\
  Plummer/Disk & $f_\star$ & - & Stellar mass fraction \\
  Plummer & $M_{\rm Pl}$ & $h^{-1}{\rm M}_\sun$ & Mass of external Plummer potential\\
  Plummer & $a$ & $h^{-1}{\rm kpc}$ & Scale of external Plummer potential\\
  Disk & $M_{\rm d}$ & $h^{-1}{\rm M}_\sun$ & Mass of external disk potential \\
  Disk & $H$ & $h^{-1}{\rm kpc}$ & Radial scale of external disk potential\\
  Disk & $z_0$ & - & Vertical scale used for SNF distribution in disk \\
  Plummer/Disk & $P$ & $h^{-1}{\rm Gyr}$ & Period of one SNF cycle \\
  Plummer/Disk & $f_{\rm g}$ & - & Fraction of SNF period over which $m(t)$ increases \\
  Plummer/Disk & $a_{\rm SNF}$ & $h^{-1}{\rm kpc}$ & Scale of individual SNF Plummer spheres\\
  Plummer/Disk & $N_{\rm SNF}$ & - & Number of SNF spheres in each period
\enddata
\end{deluxetable*}

\subsection{Spatial distribution of supernova locations}\label{sec:sndist}
In the model of \citet{2019MNRAS.485.1008B}, the SN-driven outflow is located in the centre of the halo and modelled as the sudden removal of a spherically symmetric mass distribution. Here we develop this model further by implementing a probabilistic way of assigning (fixed) positions to the superbubbles created by the outflow events across the modelled external potential (galaxy),
mimicking the fact that supernovae can occur wherever there are stars. In the following, we will occasionally refer to individual explosions within our model as ``supernovae'', implemented as outlined below. To be precise, these individual explosions are to be interpreted as approximations of superbubbles -- regions devoid of gas that are created through a spatially concentrated series of (actual) supernovae that follows a (local) episode of bursty star formation. When determining the positions of individual ``supernovae'' we make the approximation that they are more likely to occur in regions of larger stellar mass. In reality, the local density of type II supernovae is strongly correlated with the local star formation rate density. Our model cannot -- by construction -- capture local bursts in star formation. However, we note that the star formation rate is on average larger in regions where the gas is denser and since in our model ``gas'' density and ``stellar'' density have the same functional form, this means that our way of assigning ``supernova'' locations is consistent with the approximations made within our model. 
Assuming that the SN density follows the stellar density implies that if we wish to distribute SN feedback probabilistically across the mock galaxy - modeled by either an external Plummer sphere or an external disk - then we can use the (normalized) differential mass profile associated with either density profile as a probability density from which to sample the positions of individual ``supernovae''. For the Plummer sphere, this means that radii of explosion centers can be obtained from a random number $X\in (0,1)$ via
\begin{equation}
    r = \frac{a}{\sqrt{X^{-2/3}-1}}.\label{eq:random_pl}
\end{equation}
Once the radius is calculated, we determine the exact position vector by randomly selecting the angular position. In the case of a flat external disk, the enclosed mass profile cannot be analytically inverted and our task is thus to find a cylindrical radius for which 
\begin{equation}
    \frac{X-M(R)}{dM/dR} < \tau R, \label{eq:random_di_1}
\end{equation}
where $\tau$ is some numerical threshold. While we are working with the potential of a flat disk, distributing all of the ``supernovae'' exactly within the x-y plane may lead to an overestimate of the impact of SN feedback in disk galaxies. To prevent that, we assign a vertical offset to each explosion center in a probabilistic manner. To that end, we assume that in the vertical direction the mass is distributed according to a probability density $\propto \cosh(z/z_0)^{-2}$, where $z_0$ is a scale length. From a random number $X\in (0,1)$, we can then calculate the $z$-coordinate as 
\begin{equation}
    z = \frac{z_0}{2}\ln\left(\frac{X}{1-X}\right).\label{eq:random_di_2}
\end{equation}
The azimuthal angle is chosen at random from a uniform distribution. Using Equation \ref{eq:random_pl} for a Plummer sphere, and Equations \ref{eq:random_di_1} and \ref{eq:random_di_2} for a disk, we can sample any desired number of ``supernova'' locations. In the limit of a very large number of ``supernovae'', their cumulative spatial distribution will be closely related to the external potentials introduced in Section \ref{sec:plpot} (for a Plummer sphere)
and Section \ref{sec_diskpot} (for a flat disk)\footnote{In the case of the disk, we are also allowing for the possibility to introduce an additional spread in the vertical direction.}.

\subsection{Implementation of SN feedback}\label{sec:sn_imp}
In our model, ``supernovae'' occur simultaneously and with the same impact at all of the positions determined as outlined in Section \ref{sec:sndist}. At each SN location, we ``cut a hole" into the external potential, by essentially subtracting an external acceleration generated by a Plummer sphere at each SN location. The masses of these Plummer profiles are time-dependent and identical at each location. Thus, the net effect of our effective SN feedback model at a given time $t$ can be written as 
\begin{equation}
    \mathbf{a}_{\rm SNF}(\mathbf{r}) = \sum_{i=1}^{N_{\rm SNF}}\frac{Gm(t)}{\left({a^2_{\rm SNF}+(\mathbf{r}-\mathbf{r}_{{\rm SNF},i})^2}\right)^{3/2}}(\mathbf{r}-\mathbf{r}_{{\rm SNF},i}).\label{eq:net_snfeedback}
\end{equation}
where $\mathbf{r}_{{\rm SNF},i}$ denotes the position vector of the {\it i}'th SN, $a_{\rm SNF}$ is a Plummer scale that prevents exceedingly short time-steps for particles that get too close to a particular SN and $m(t)$ is the time-dependent mass of the Plummer profile used to model SN feedback. In our model, we increase $m(t)$ linearly over a fixed amount of time and then decrease it linearly over a longer time. This process happens periodically throughout the simulation and for each new period new SN positions are sampled as described in Section \ref{sec:sndist}. The energy release period $P$ is a free input parameter, as well as the fraction of each period during which $m(t)$ increases until it reaches its maximum value ($f_g$). The number of ``supernovae'' during each period, $N_{\rm SNF}$, is a further input parameter of our model. The positions of individual ``supernovae'' are fixed during the energy release period $P$. This constitutes a simplifying approximation since the positions of actual superbubbles are time-dependent -- due to the streaming motion of the surrounding gas. In principle, assuming static ``supernova'' locations may introduce an additional degree of asymmetry into the system, an effect that we aim to minimize by choosing a sufficiently large number of ``supernovae'', $N_{\rm snf}$, during each explosion cycle.

To connect the energy that is injected through our SN feedback model to the total stellar mass modeled by the external Plummer/disk potential, we attempt to connect the maximum value of $m(t)$ to the model parameters in a self-consistent way. 
To estimate the amount of energy that is injected into the interstellar medium through SN explosions by a galaxy of stellar mass $M_\star$ we follow Equation 6 of \citet{Penarrubia:2012bb},
\begin{equation}
    \Delta E = \frac{M_\star}{\left<m_\star\right>}\xi(m_\star > 8{\rm M}_\sun)E_{\rm SN}\epsilon. \label{eq:penarrubia}
\end{equation}
Equation \ref{eq:penarrubia} renders the available SN feedback energy as a function of the mean stellar mass $\left<m_\star\right>$, the fraction of stars with a mass larger than $8{\rm M}_\sun$, $\xi(m_\star > 8{\rm M}_\sun)$, the typical energy of one supernova, $E_{\rm SN}$, and the effective coupling efficiency of SN feedback to the interstellar medium, $\epsilon$. In our model, the coupling efficiency is a free parameter. For the other parameters in Equation \ref{eq:penarrubia}, we follow \citet{Penarrubia:2012bb} and set $\left<m_\star\right> = 0.4{\rm M}_\sun$, $\xi(m_\star > 8{\rm M}_\sun) = 0.0037$, and $E_{\rm SN} = 10^{51}{\rm erg}$, where the first two values originate from using a \citet{2002Sci...295...82K} initial mass function, while $E_{\rm SN}$ is the canonical kinetic energy released in SN type II explosions \citep[e.g.][]{refId0}.

Equation \ref{eq:penarrubia} estimates the total energy that is returned to the interstellar medium (ISM) by SN feedback. The total increase in the virial energy of the DM halo determines how large the eventual formed core can be (see \citealt{Penarrubia:2012bb}). However, how much of the energy that is injected into the interstellar medium couples to the DM is still uncertain. In principle, the coupling efficiency between the energy injected by supernovae and the DM, $\epsilon_{\rm DM}$, depends on many factors. Since this coupling is purely gravitational, it can depend on the positions of individual superbubbles, their lifetime, and the local DM density. Since we can measure the energy of individual DM particles throughout the simulated time, we can approximately determine how $\epsilon_{\rm DM}$ depends on the various settings of our model. However, to choose sensible values for the parameters of our effective SN feedback model, we need to obtain a reasonable a priori guess for the total energy that is injected into the interstellar medium (Equation \ref{eq:penarrubia}) by SN feedback. To that end, we assume that the stellar mass $M_\star$ within the external Plummer (disk) baryonic potential is given by $M_\star = f_\star M_{\rm Pl}$ ($M_\star = f_\star M_{\rm d}$). If there are $N_{\rm P}$ SN feedback periods (cycles) during our simulation, the injected energy during each period is $\Delta E_{\rm P} = \Delta E / N_{\rm P}$. 
The energy associated with each individual ``superbubble'' is then equal to $E_{\rm SB} = \Delta E_{\rm P}/N_{\rm SNF}$. We then identify this energy with the gravitational binding energy of one of the Plummer spheres that we use to model superbubbles, and use this correspondence to fix the maximum removed ``mass'': 
\begin{equation}
    E_{\rm SB} = \frac{3\pi}{32}\frac{Gm_{\rm max}^2}{a_{\rm SNF}}.\label{eq:sn_selfen}
\end{equation}
We note that Equation \ref{eq:sn_selfen} is an approximation that can only provide a rough order of magnitude estimate for the energy injected by each one of the ``supernovae''. The central assumption is that the energy of the ``supernovae'' is equal to the gravitational binding energy of the removed baryonic material of mass $m_{\rm max}$, distributed following a Plummer profile as implemented in our model. For this approximation to be (at least approximately) applicable, the density associated with the removed material needs to be significantly larger than the baryonic density in the surroundings. If this is not the case, the gravitational interaction between the removed material and the surrounding mass contributes significantly to the total energy budget, and the total injected energy will be significantly larger than the nominal energy quoted in Equation \ref{eq:penarrubia}. A way to achieve that
Equation \ref{eq:sn_selfen} 
approximates the injected energy is %
if the local gas density within the (fully formed) ``superbubbles'' is negative. While negative gas densities are clearly unphysical, they are not a problem within our model as long as the surrounding DM particles remain gravitationally bound at the end of each supernova cycle -- otherwise the DM halo would be artificially disrupted. It is thus advisable to choose $N_{\rm SNF}$ in such a way that the local gas density within fully formed ``superbubbles'' is negative without unbinding the neighbouring DM particles. 
Once all of the model parameters are set, we use Equation \ref{eq:sn_selfen} to calculate the maximal mass $m_{\rm max}$ of the individual Plummer spheres that mimic localized starbursts/outflows across the modeled galaxy. The 
mass $m(t)$ defined in Equation \ref{eq:net_snfeedback} is then given by
\begin{equation}
    m(t) = \left\{ \begin{array}{ll}
        m_{\rm max}\frac{t}{f_{\rm g}P} & t \le f_{\rm g}P \\
        m_{\rm max}\frac{P-t}{P(1-f_{\rm g})} & t > f_{\rm g}P
    \end{array}  \right.,
\end{equation}
where $t$ is the simulation time modulo the period $P$. 
Having fixed the external potential of the Plummer (disk) galaxy, as well as the SN feedback associated with either of those potentials, our effective model for SN feedback is now almost fully defined. An explanation of how we determine the number of SN feedback periods from a simulation's parameter file will follow when we discuss the setups of our simulations. Moreover, we discuss the energy that is actually injected into the DM halo, and how it compares to the energy that is nominally injected into the ISM (defined as outlined above), in Section \ref{sec:reseng}.

A summary of the parameters defining the effective model for SN feedback, along with a brief description of the role of each parameter, can be found in Table \ref{tab:model_params}.

\section{Simulations} \label{sec:simulations}
Through a series of controlled simulations of an isolated dwarf-sized DM halo, we aim to both test our effective model for SN feedback and investigate the impact of varying three key parameters: 
$\epsilon$, $f_{\rm g }$, as well as $a$ (in the Plummer case) or $H$ (in the disk case). 
In this Section, we outline the steps taken to conduct this series of simulations. First, we discuss how we obtain initial conditions of a halo in approximate dynamical equilibrium. Then, we outline how we set up orbital families of tracer particles (akin to how it was done in \citealt{2019MNRAS.485.1008B}) 
to track whether changes in the halo's gravitational potential are adiabatic or impulsive. Finally, we discuss the adopted model parameters (see table \ref{tab:model_params}) for each of the simulations in our simulation suite.  

\subsection{Initial conditions}\label{sec:ICS}
For each of our initial conditions, we start by self-consistently sampling a live halo of collision-less DM. We use \citet{Eddington1916}'s formalism to construct a distribution function for a NFW halo \citep{1996ApJ...462..563N} with a DM mass of $M_{200} = 10^{10}h^{-1}{\rm M}_\sun$ and an initial concentration of $c_{200} = 13$, where $c_{200} = r_{200}/r_s$, i.e., the ratio between the halo's virial radius and its scale radius; for definiteness we refer to virial quantities for the halo properties corresponding to a virial radius $r_{200}$ enclosing an average density equal to 200 times the critical density of the Universe today. Beyond $r_{200}$, we exponentially cut off the density profile in order to avoid an infinitely massive halo. 
Having constructed the distribution function, we draw radii from the analytic differential mass profile of such a halo and then use a rejection sampling method to self-consistently sample the DM particles' velocities from the distribution function (assuming an isotropic velocity dispersion tensor). 
\begin{figure*}
    \centering
    \includegraphics[width=0.49\linewidth]{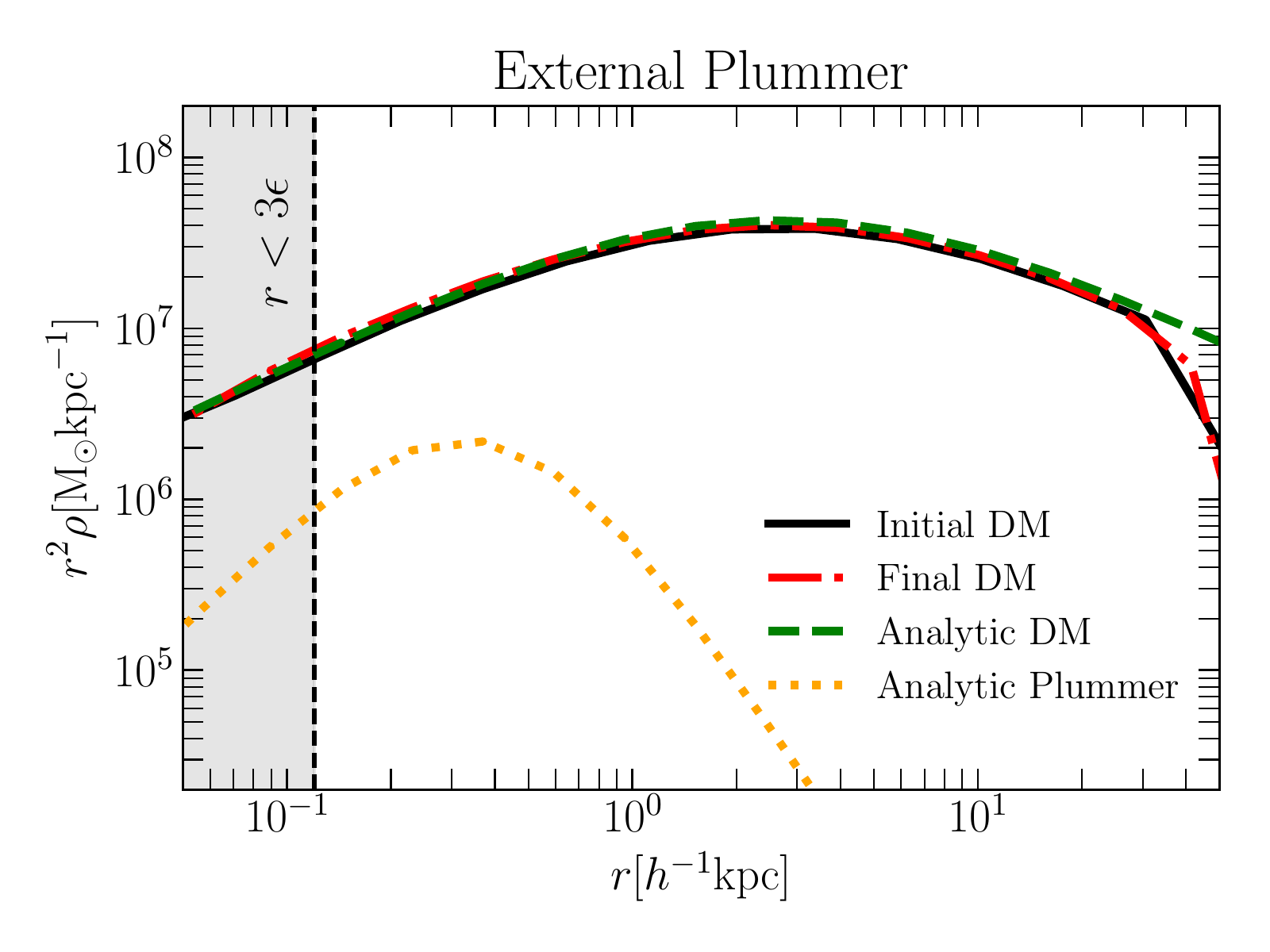}
    \includegraphics[width=0.49\linewidth]{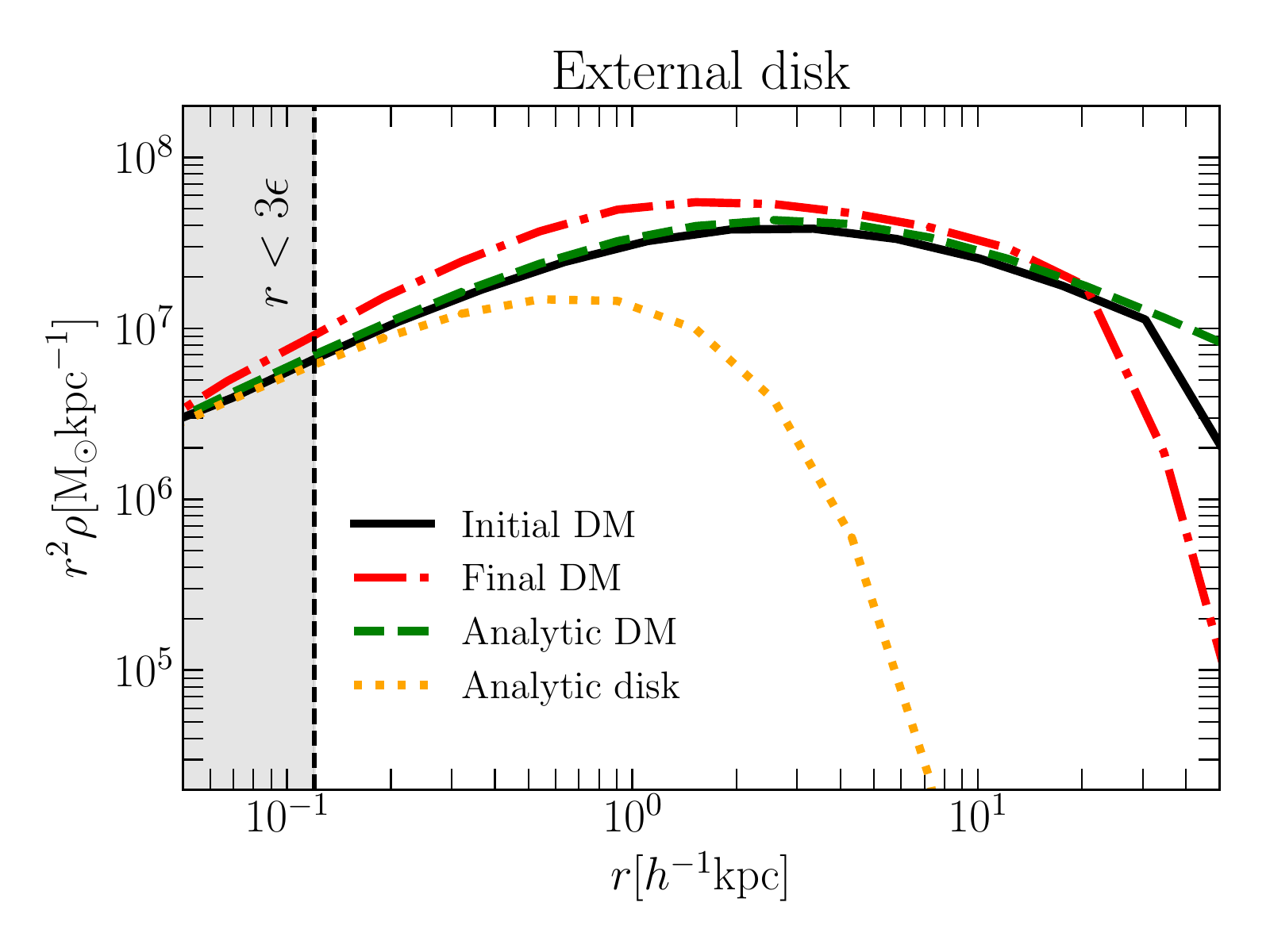}
    \caption{Construction of the (equilibrium) initial conditions for our simulations, exemplified on the halo including a benchmark external Plummer (disk) potential on the left (right) panel. The grey shaded area indicates the region affected by numerical resolution. The black solid lines show the initial density profile of the simulated DM particles, calculated in spherical shells around the halo's center of potential. This profile overlaps with an analytic NFW profile (green dashed lines) for radii smaller than $r_{200}$. The dotted orange lines show the (spherically averaged) density profile equivalent to adding the external Plummer (disk) potential to the simulation. The red dash-dotted lines show the contracted DM profiles after relaxing for $1h^{-1}{\rm Gyr}$ of simulation time. These final profiles represent equilibrium configurations that are used as initial conditions for subsequent simulations. To highlight the effect of the adiabatic contraction, all profiles are multiplied by $r^2$.}
    \label{fig:ICgeneration}
\end{figure*}

After self-consistently sampling position and velocity vectors of $10^7$ DM particles, we obtain an $N$-body representation of an isolated halo in approximate dynamical equilibrium\footnote{The equilibrium can never be perfect due to the softened force law used in collisionless $N$-body simulations.}. However, adding an external -- Plummer or disk -- potential will put the system out of dynamical equilibrium. In order to start our simulations from controlled initial conditions, we thus let the halo (plus external potential) evolve until the system settles into a new state of dynamical equilibrium, using the code {\scriptsize AREPO} \citep{Springel:2009aa} to calculate the self-gravity between the DM particles. 

Since we want to investigate the impact of varying the concentration of the galaxy, we perform several preparation runs -- one for each variation of the external potential. For illustration, we chose external potentials that are inspired by the properties of observed MW satellites. In particular, we chose a benchmark setup for both the external Plummer potential and the external exponential disk. The parameters of the benchmark Plummer potential are inspired by observed properties of Fornax. Following \citet{2012AJ....144....4M}, Fornax has a stellar mass of $2\times 10^7{\rm M}_\sun$ and a half-light radius of $r_{1/2} = 0.7\,{\rm kpc}$. Moreover, \citet{2016MNRAS.456.3253Y} show that while gas in Fornax is fully depleted today, the gas mass in Fornax has been on the order of or slightly larger than its stellar mass for a large part of its evolutionary history. With our benchmark Plummer model, we approximate a Fornax-like dwarf by choosing the parameters $M_{\rm Pl} = 2\times 10^7 h^{-1}{\rm M}_\sun$, $f_\star = 0.5$, and $a = 0.4\,h^{-1}{\rm kpc}$ (see Table \ref{tab:model_params} for an explanation). To investigate the effect of a more (less) compact galaxy, we halve (double) the scale radius $a$. With our benchmark disk model, we aim to roughly capture the gravitational effect of the Small Magellanic Cloud (SMC). \citet{2012MNRAS.421.3488H} models an SMC-like galaxy as a disk with a total baryonic mass of $M_{\rm d} = 8.9\times 10^8{\rm M}_\sun$. The stellar disk has a scale length of $H = 0.7\,{\rm kpc}$. However, the gaseous component extends much further outwards and is the dominant component in terms of mass (see Table 1 of \citealt{2012MNRAS.421.3488H}) while our model does not allow for two different disks with different scale lengths. Moreover, \citet{2012MNRAS.421.3488H} assumes a host halo mass which is significantly larger than the mass of our fiducial DM halo. For these reasons, we chose to use slightly different parameters in our benchmark SMC-like disk: 
$M_{\rm d} = 4\times 10^8 h^{-1}{\rm M}_\sun$, $f_\star = 0.15$, and $H = 0.7\,h^{-1}{\rm kpc}$. Aside from the benchmark model, we also investigate the effect of having a more (less) compact disk by halving (doubling) the disk's scale length. All external potentials are static, meaning that they are added instantaneously at the beginning of the preparation runs. 

The initial conditions for our eventual simulations are the final snapshots of the preparation runs conducted for a total time of $1h^{-1}{\rm Gyr}$\footnote{For a NFW profile with $c_{200} = 13$, this corresponds to the orbital period of a particle on a circular orbit with radius $r_{\rm circ} \approx 7\,h^{-1}$kpc.}, in which the DM particles have had the time to respond to the addition of the external potential and settle into a new dynamical equilibrium. 

In Figure \ref{fig:ICgeneration}, we show the process of generating the initial conditions of our simulations, exemplified on the two benchmark cases (Plummer on the left panel, disk on the right panel). The grey shaded area indicates the radial range in which the measured density profiles are not fully reliable, i.e. radii which are smaller than three times the gravitational softening length of the simulation \citep{Power:2002sw}. As coloured lines we show the density profiles corresponding to the external potentials, as well as the DM density profiles at different times, scaled by $r^2$. The green dashed lines and the black solid lines are the same across both panels. In green-dashed, we show an analytic NFW profile with the target virial mass and concentration. 
In black, we show the halo's initial DM density profile -- before adding the external potential. Over a large range of radii those two lines coincide, but at radii larger than $r_{200}$ we see the exponential cutoff included to have 
a numerical solution to Eddington's equation. The spherically averaged density profile corresponding to the added external Plummer (disk) potential is shown as an orange dotted line on the left (right) panel. Finally, the red dash-dotted lines show the DM density profiles measured after letting the system (DM halo plus external galactic potentials) relax for $1h^{-1}{\rm Gyr}$. When adding the external Plummer potential, we find that the final and the initial DM density profiles are almost exactly identical: the relatively small additional mass causes no significant contraction in the DM. In the case of an external disk potential, however, the mass of the galaxy causes a significant contraction in the DM density profile. We thus anticipate 
that the gravitational pull of the external disk can be of significant importance when it comes to whether (and how quickly) the DM cusp can be restored after impulsive removal of baryonic mass from the halo's center. 
To ensure that the final, adiabatically contracted DM density profile is fully determined by the structural parameters of the external potential -- and does not depend on the rate at which the external potential is added -- we have conducted one additional preparation run. In this run, instead of instantaneously adding the benchmark external disk potential, the ``mass'' of the external disk is linearly increased over $600\,h^{-1}$Myr, before we let the system relax for another $400\,h^{-1}$Myr. We found no obvious differences between the final DM density profile in this additional preparation run and the final DM density profile shown on the right panel of Figure \ref{fig:ICgeneration}.


\subsection{Orbital Families and Explosion Times}\label{sec:ofamdef}

In order to track whether the sizes of cores formed through SN feedback correlate with how impulsive our implementation of SN feedback is, we set up orbital families as in \citet{2019MNRAS.485.1008B} to investigate how they respond to the changes in potential. Depending on whether the orbital family remains united or splits into several families of orbits, we can tell whether the implemented SN feedback was adiabatic or impulsive. 

To the initial conditions described in Section \ref{sec:ICS} for the Plummer spheres,
we add a family of orbits as described in Section 4.1 of \citet{2019MNRAS.485.1008B}. We sample 2000 tracers with pericenter radii of $r_{\rm peri} = 0.5\pm 0.05\, h^{-1}{\rm kpc}$ and apocenter radii of $r_{\rm apo} = 2\pm 0.05\, h^{-1}{\rm kpc}$. Note that in a spherically symmetric potential, this is equivalent to sampling orbits with similar energies and angular momenta. Since in a fixed spherically symmetric potential the radial action is only a function of energy and angular momentum, asking whether or not an orbital family of tracers remains united (and thus whether the SN feedback is adiabatic or impulsive) is equivalent to asking whether or not radial actions are conserved. 

Setting up the orbital family works slightly differently for the initial conditions corresponding to the three external disk potentials. Since these potentials are axisymmetric, 
we have to restrict the orbits of the tracers 
to be within the plane of the disk. Hence, instead of $\Phi(r)$ we now consider the in-plane potential $\Phi(R,0)$, and we initialize all tracers in-plane and with no vertical velocity component ($z = 0, v_z = 0$). 
We note, however, that a slight deviation from cylindrical symmetry throughout the simulation can cause perturbations to the orbits that may cause the orbital families to diffuse, irrespective of whether these perturbations occur on adiabatic or impulsive timescales. With that in mind, we will see if signatures of impulsive SN feedback will still be apparent in the case of an external disk potential, but we do not require that the orbital family stay fully united in order to classify a particular setup of the effective model for SN feedback as adiabatic.   

In our simulation suite, we also aim to investigate the impact of changing the time over which the energy from SN feedback is injected into the DM distribution. To enable a comparison between the 
different external potentials, we express the injection times as fixed fractions of the radial period of a particle which is part of the orbital family. The radial periods $\mathcal{P}_r$ for the Plummer potentials and $\mathcal{P}_R$ for the disk potentials are 
\begin{equation}
    \mathcal{P}_r = \int_{r_{\rm peri}}^{r_{\rm apo}}\frac{dr}{v_r} \qquad \mathcal{P}_R = \int_{R_{\rm peri}}^{R_{\rm apo}}\frac{dR}{v_R}.
\end{equation}
Setting up our simulations, we then chose growth fractions $f_{\rm g}$ such that the mass $m(t)$ grows as outlined in Section \ref{sec:sn_imp} for a time which equals 1 per cent, 10 per cent, or 200 per cent of the radial period of a particle with pericenter radius $r_{\rm peri} = 0.5\,h^{-1}{\rm kpc}$ and apocenter radius $r_{\rm apo} = 2\,h^{-1}{\rm kpc}$.

\subsection{Simulation Settings}\label{ec:simset}

\begin{deluxetable}{lcCCCC}
\tablenum{2}
\tablecaption{Parameters that vary between simulations: 
nominal energy coupling to the ISM ($\epsilon$, column 3), fraction of explosion cycle over which the energy is injected ($f_{\rm g}$, column 4) and concentration of the external potential ($a$ for the Plummer sphere, column 5, and $H$ for the exponential disk, column 6). For each parameter, we investigate a benchmark value, as well as a value that should cause a more adiabatic (impulsive) SN feedback. \label{tab:simu_params}}
\tablewidth{0pt}
\tablehead{
\colhead{Galaxy} & \colhead{Parameter} & \colhead{$\epsilon$} & \colhead{$f_{\rm g}$} & \colhead{$a$} & \colhead{$H$} \\
\colhead{type} & \colhead{impact} & & & \colhead{$h^{-1}{\rm kpc}$} & \colhead{$h^{-1}{\rm kpc}$}
}
\decimalcolnumbers
\startdata
    & adiabatic & 0.01 &  0.33 & 0.8 & - \\
  Plummer & benchmark & 0.05 & 0.017 & 0.4 & - \\ 
   & impulsive & 0.4 & 0.0017 & 0.2 &- \\
   \hline
   & adiabatic & 0.01 & 0.46 & - & 1.4  \\
  Disk & benchmark & 0.05 & 0.023 & - & 0.7\\
   & impulsive & 0.4 & 0.0023 & - & 0.35\\
\enddata
\end{deluxetable}

Having generated all the required initial conditions, we now look to run a simulation suite in order to test the effective model for SN feedback and to investigate how changing the total energy input, the injection time, and the concentration of the external potentials affects the final DM density profiles, as well as the orbital family. Table \ref{tab:simu_params} shows the different numerical values that we adopt for the parameters regulating energy input, injection time, and concentration. For each parameter, we define one benchmark value, as well as one value which should make SN feedback more adiabatic and one value which should make SN feedback more impulsive. We investigate each possible combination of these parameters in order to determine which one of them causes larger cores in the DM profiles and whether core size relates directly to how adiabatic / impulsive the change in gravitational potential induced by our effective model for SN feedback is. This means that in total we run $3^3 = 27$ simulations for each of the two galaxy-like potentials.

All of the other model parameters (see Table \ref{tab:model_params}) are fixed to benchmark values in all simulations. Let us here briefly introduce and discuss their numerical values:
\begin{itemize}
    \item $M_{\rm Pl}=2\times 10^7\,h^{-1}{\rm M}_\sun$ and $M_{\rm d}=4\times 10^8\,h^{-1}{\rm M}_\sun$ as discussed in Section \ref{sec:ICS}.
    \item $f_\star$ is 0.5 and 0.15 for the Plummer sphere and disk, respectively. 
    This choice is to mimic the effect of a Fornax-like dwarf in the former case and an SMC-like galaxy in the latter case (see  Section \ref{sec:ICS}). 
    \item $a_{\rm SNF}$ is set to $10\,h^{-1}{\rm pc}$. This sets the scale of individual ``superbubbles'' to be smaller than the gravitational softening in our simulation, assuring that the energy injection is effectively point-like and thus that the local density contrast introduced by the ``superbubbles'' is large (see relevant discussion in Section \ref{sec:sn_imp}). 
    \item $z_0$ is always determined as $0.2 H$, in agreement with \citet{2012MNRAS.421.3488H}.
    \item For definiteness, $P$ is set to $0.6h^{-1}{\rm Gyr}$ in all simulations. From the period, we can fix the number of explosion periods $N_P$ in all simulations. Before the first wave of explosions, we wait for one period in order to monitor whether the orbital family remains united in the absence of SN feedback. Additionally, we want the system to relax at the end of the simulations, in order for the final DM density profiles to not be affected directly by the gravitational impact of the explosion centers. To that end, no explosions are implemented during the last $1\,h^{-1}{\rm Gyr}$ of the simulations. We run all simulations for a total time of $4\,h^{-1}{\rm Gyr}$, which implies that the number of explosion cycles is $N_P = 3/0.6 -1 = 4$. SN feedback is thus distributed over four explosion cycles of equal duration, with $N_{\rm SNF}=20$ explosion centers during each cycle. 
\end{itemize}

As stated in Section \ref{sec:ICS}, we use the {\scriptsize AREPO} code to determine the self-gravity and the time evolution of the isolated systems. For collisionless particles, {\scriptsize AREPO} uses adaptive time-steps. The adopted time-step criterion is based on the softening lengths of individual particles (see Equation 34 of \citealt{Springel:2005mi}). Gravitational forces between different simulation particles are calculated using a hierarchical multipole expansion. 
A relative cell opening criterion is used (see Equation 18 in \citealt{Springel:2005mi}). In all our simulations and preparation runs, we choose softening lengths of $40\,h^{-1}$pc ($1h^{-1}$pc) for the DM (tracer) particles and an accuracy parameter for the cell opening criterion $\alpha = 0.0005$.

The calculation of each particle's time-step is based on its total acceleration, taking into account both self-gravity and the external forces generated by our effective model of supernova feedback (see Section \ref{sec:method}). In our model, the external accelerations can change rapidly over a short time and as a consequence, the time-steps of particles that are near a ``supernova location'' may occasionally be too long just before the ``explosion", and thus, their acceleration may not be updated fast enough. To verify that this does not significantly affect our results, we have repeated one of our simulations (the Plummer run with $\epsilon = 0.05$, $a = 0.2\,h^{-1}$kpc, and $f_g = 0.017$), but this time fixing the time-step of all DM particles to the minimum value reached in the run with adaptive time-steps. We have found that both the size of the final core and the time evolution of the DM density profile are in good agreement between the two simulations. 

\begin{figure*}
    \centering
    \includegraphics[width=0.47\linewidth]{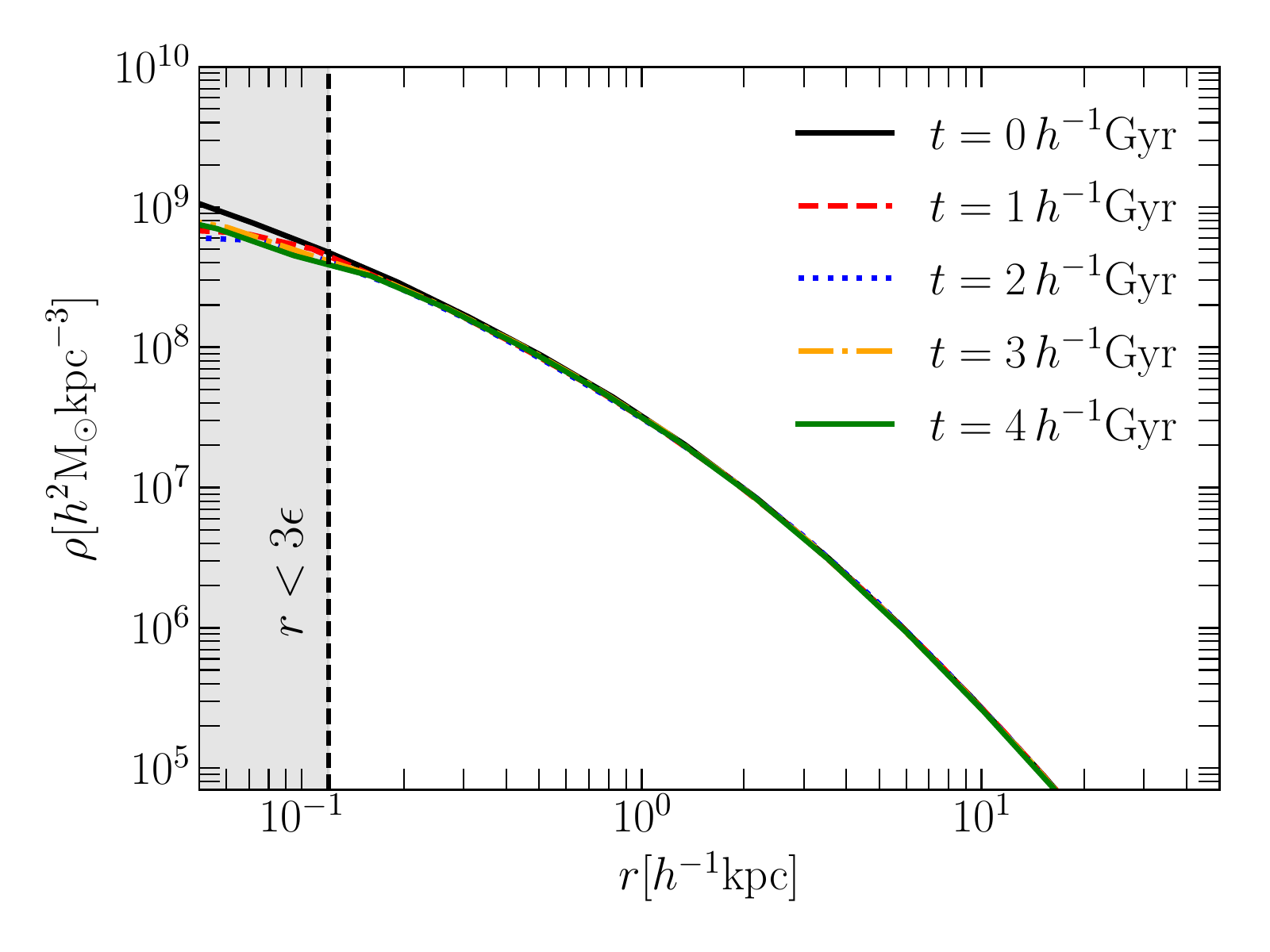}
    \includegraphics[width=0.52\linewidth]{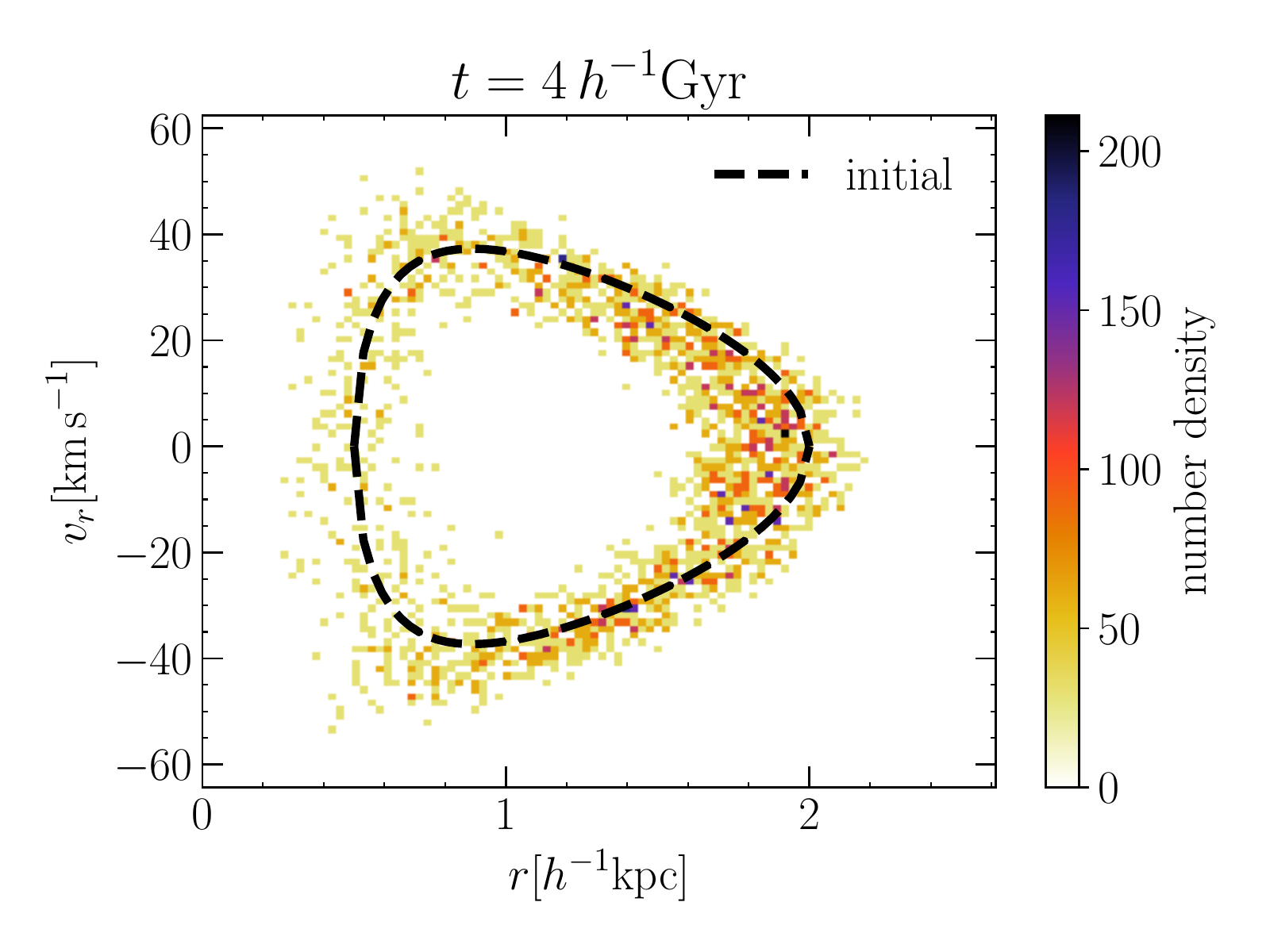} \\
    \includegraphics[width=0.47\linewidth]{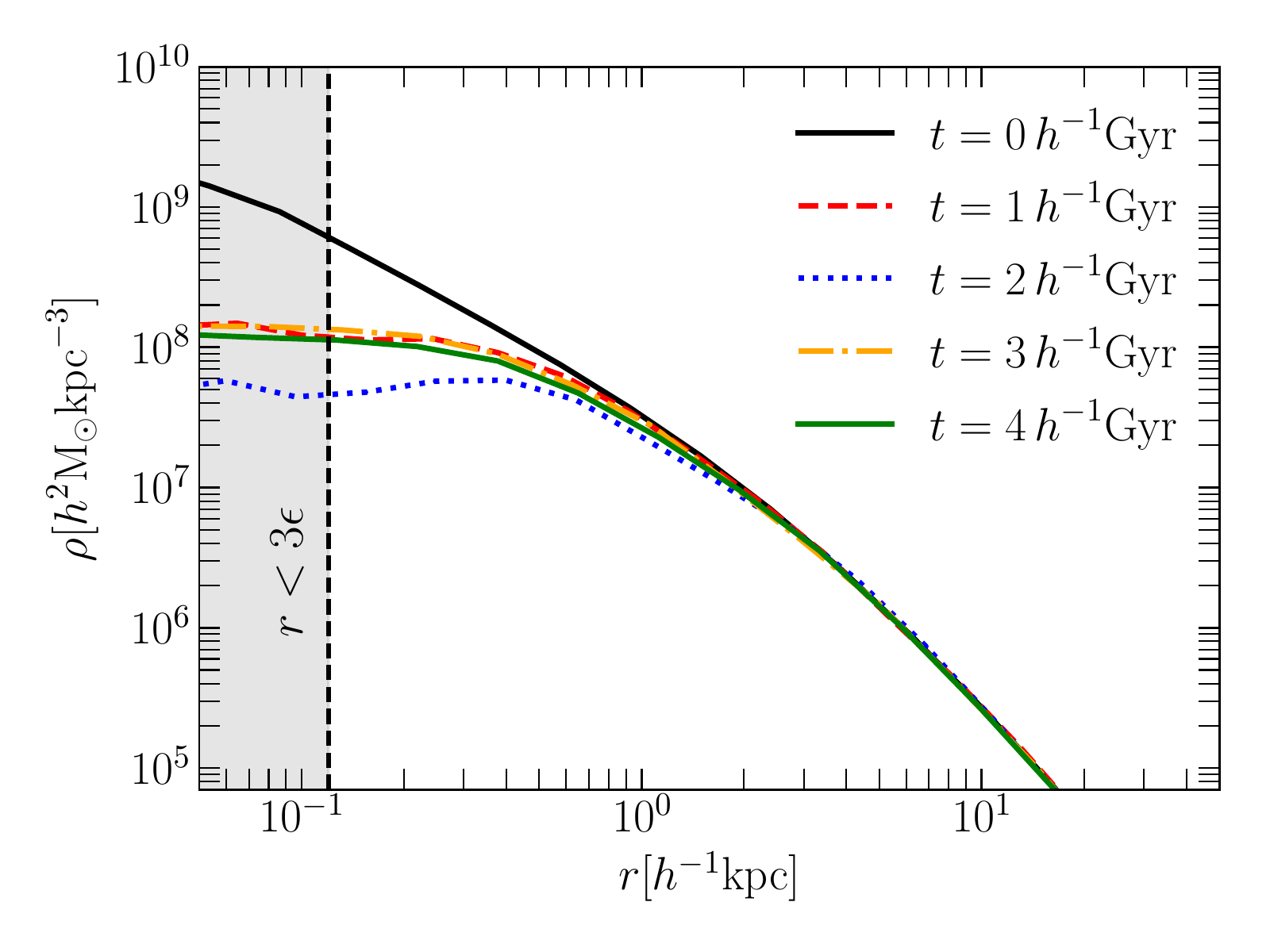}
    \includegraphics[width=0.52\linewidth]{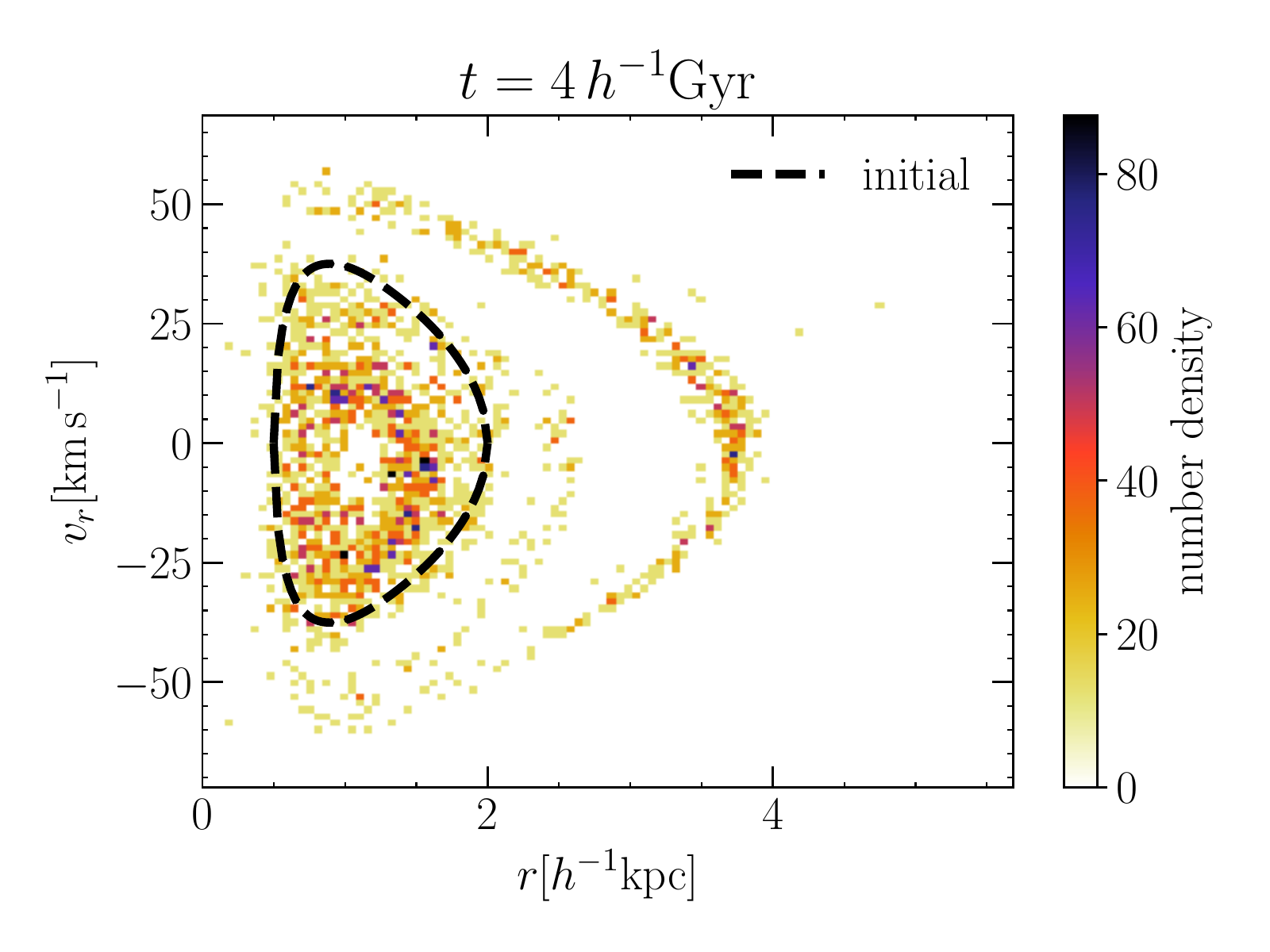}
    \caption{The results of two DMO simulations including an external Plummer sphere potential are shown. In the left column, we show the measured DM density profiles at different simulation times in intervals of $1\,h^{-1}{\rm Gyr}$. The grey shaded area indicates the region affected by numerical resolution. 
    In the right column, we show the final phase space density distribution of tracer particles that were initially set up as an orbital family. The black dashed lines show the initial ``central'' phase space trajectory of the orbital family. 
    The upper panels correspond to the simulation in which $\epsilon = 0.01$ (adiabatic), $f_{\rm g} = 0.33$ (adiabatic), and $a = 0.8\,h^{-1}{\rm kpc}$ (adiabatic). The lower panels correspond to the simulation with $\epsilon = 0.05$ (benchmark), $f_{\rm g} = 0.017$ (benchmark), and $a = 0.2\,h^{-1}{\rm kpc}$ (impulsive).}
    \label{fig:Pl_feedback}
\end{figure*}

\section{Results}\label{sec:results}
First, we present examples of runs in which a core has formed and compare them to cases in which the 
density profile remains cuspy. Subsequently, we discuss how the symmetry of the system affects the kinematics of the orbital family of tracers. Then, we compare the final profiles of all simulations 
to discuss the impact of changing the nominal energy coupling, the injection time and the concentration of the external potentials. Thereafter, we compare the effective change in energy of the DM particles to the nominally injected energy (see Equations \ref{eq:penarrubia} and \ref{eq:sn_selfen}). Based on this comparison, we discuss the accuracy of our a priori guess for the amount of energy that is injected into the ISM (based on Equation \ref{eq:sn_selfen}), as well as how the (gravitational) coupling of the injected energy to the DM depends on the energy injection time and the size and shape of the external galaxy. Finally, we briefly discuss whether impulsive energy injection is a necessary condition for core formation through SN feedback. 

\subsection{Cored vs. cuspy profiles}\label{sec:corevscusp}
Here 
we compare the evolution of the DM density profiles, as well as the final phase space distribution of the orbital family, between a simulation in which the DM halo forms a core and a simulation in which it retains its cusp. We focus separately on the cases of an external Plummer sphere and exponential disk. 

\subsubsection{Feedback from a Plummer sphere}\label{sec:fbplummer}

In Figure \ref{fig:Pl_feedback}, we compare the results of two different simulations including an external Plummer potential, with differently regulated SN feedback. The upper two panels show results of the simulation in which all of the parameters introduced in Table \ref{tab:simu_params} lean towards an adiabatic change in the potential. The left panel shows the evolution of the DM density profile 
after each $1\,h^{-1}{\rm Gyr}$ of simulation time. The grey shaded area denotes the range in which the initial DM profile cannot be considered stable according to the \citet{Power:2002sw} stability 
criterion. For radii that lie outside this range, however, we hardly detect any evolution in the DM density profile. The halo retains its cusp with our SN feedback model having no significant impact on the DM distribution. 
In agreement with that, the final phase space distribution of the orbital family of tracers 
(upper right panel) remains united by the end of the simulation. Relative to the initial phase space distribution -- which is distributed very closely around the black dashed line -- 
we detect a non-negligible diffusion of the orbital family, which is due to the fact that the gravitational potential is in fact changing (slowly) with time. However, we see no signatures of an impulsive change in the gravitational potential, i.e., neither does the orbital family split up, nor do the orbits expand to larger radii on average (see \citealt{2019MNRAS.485.1008B} for a more detailed analysis).  

\begin{figure*}
    \centering
    \includegraphics[width=0.47\linewidth]{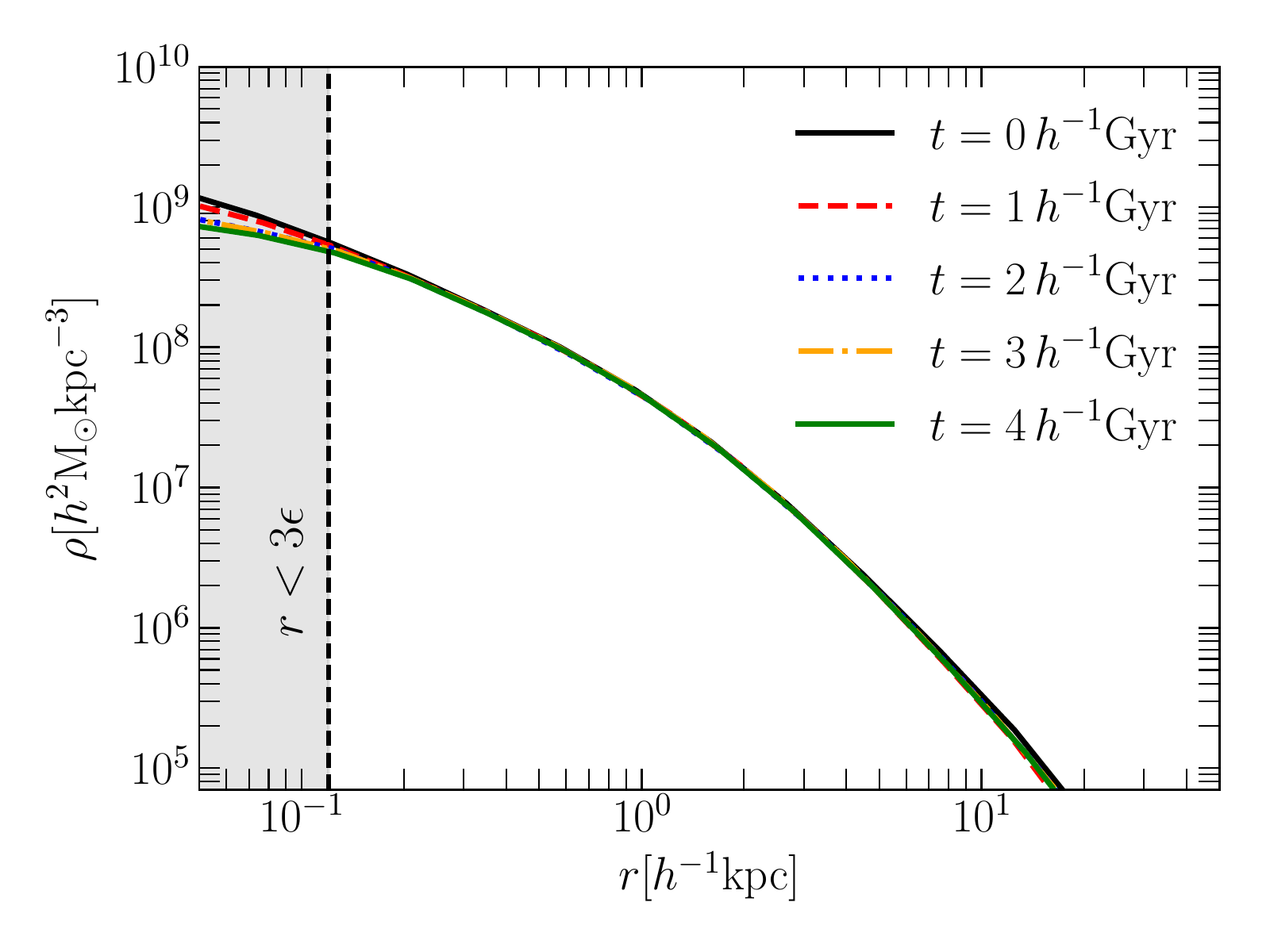}
    \includegraphics[width=0.52\linewidth]{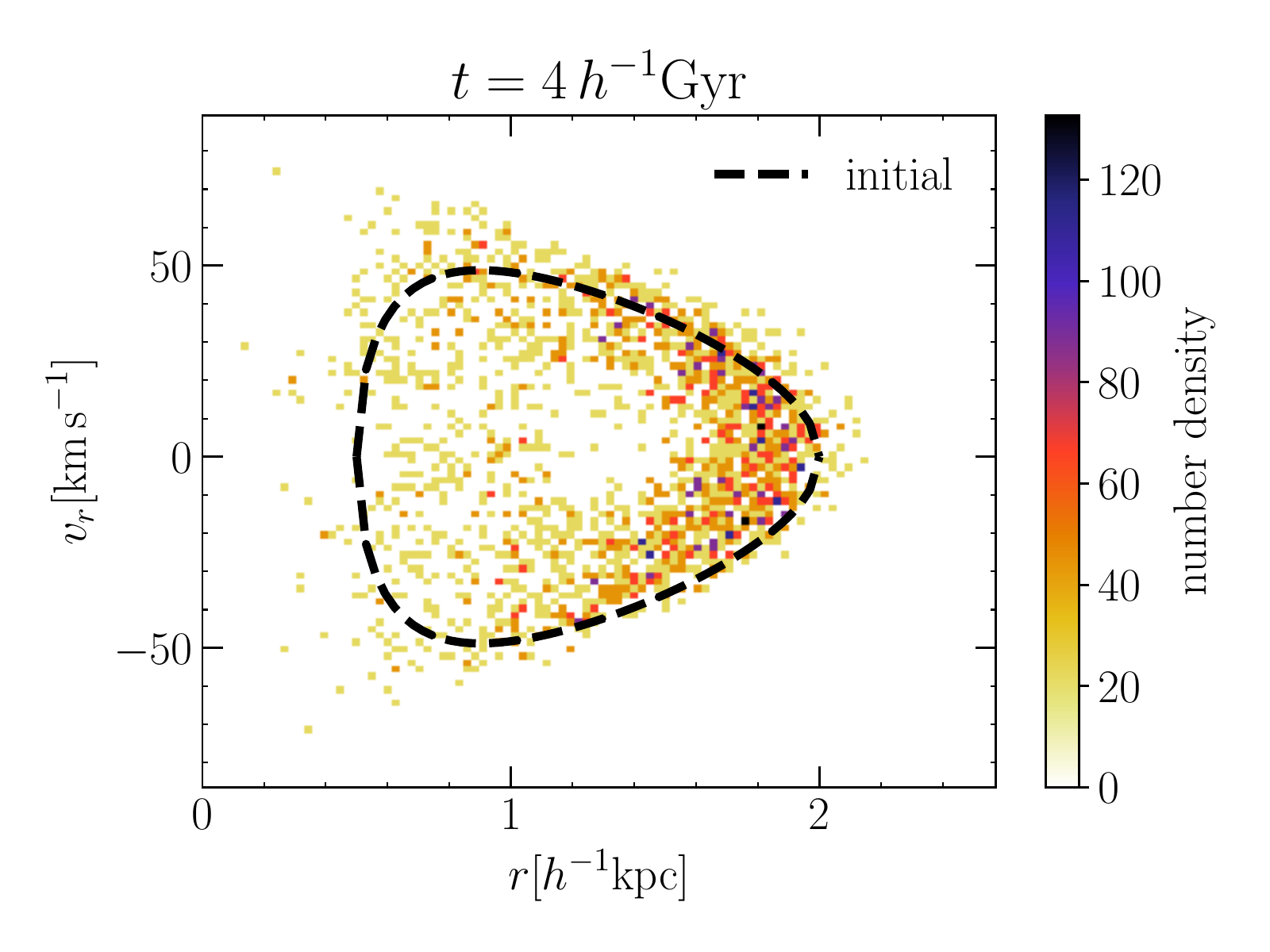} \\
    \includegraphics[width=0.47\linewidth]{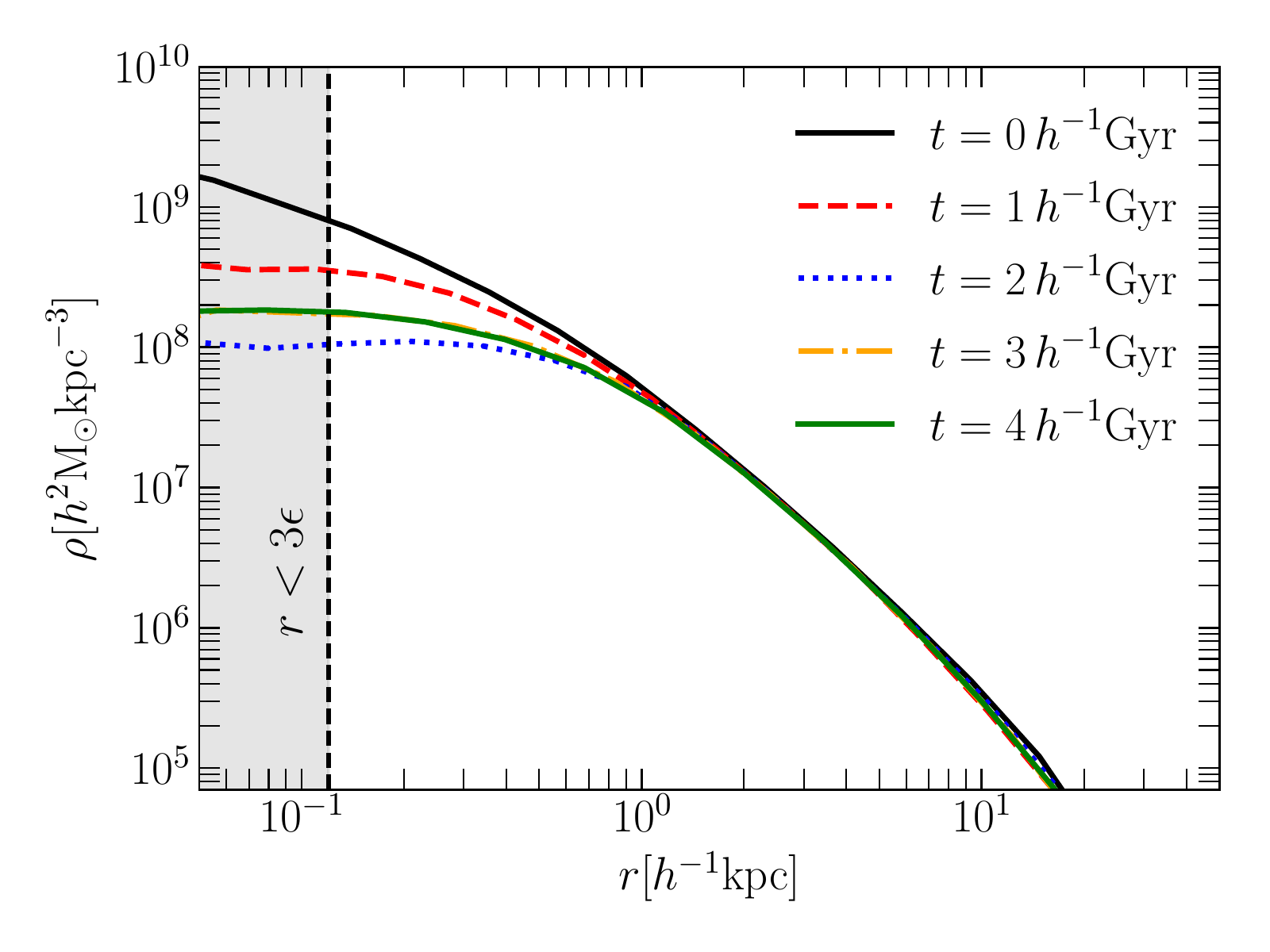}
    \includegraphics[width=0.52\linewidth]{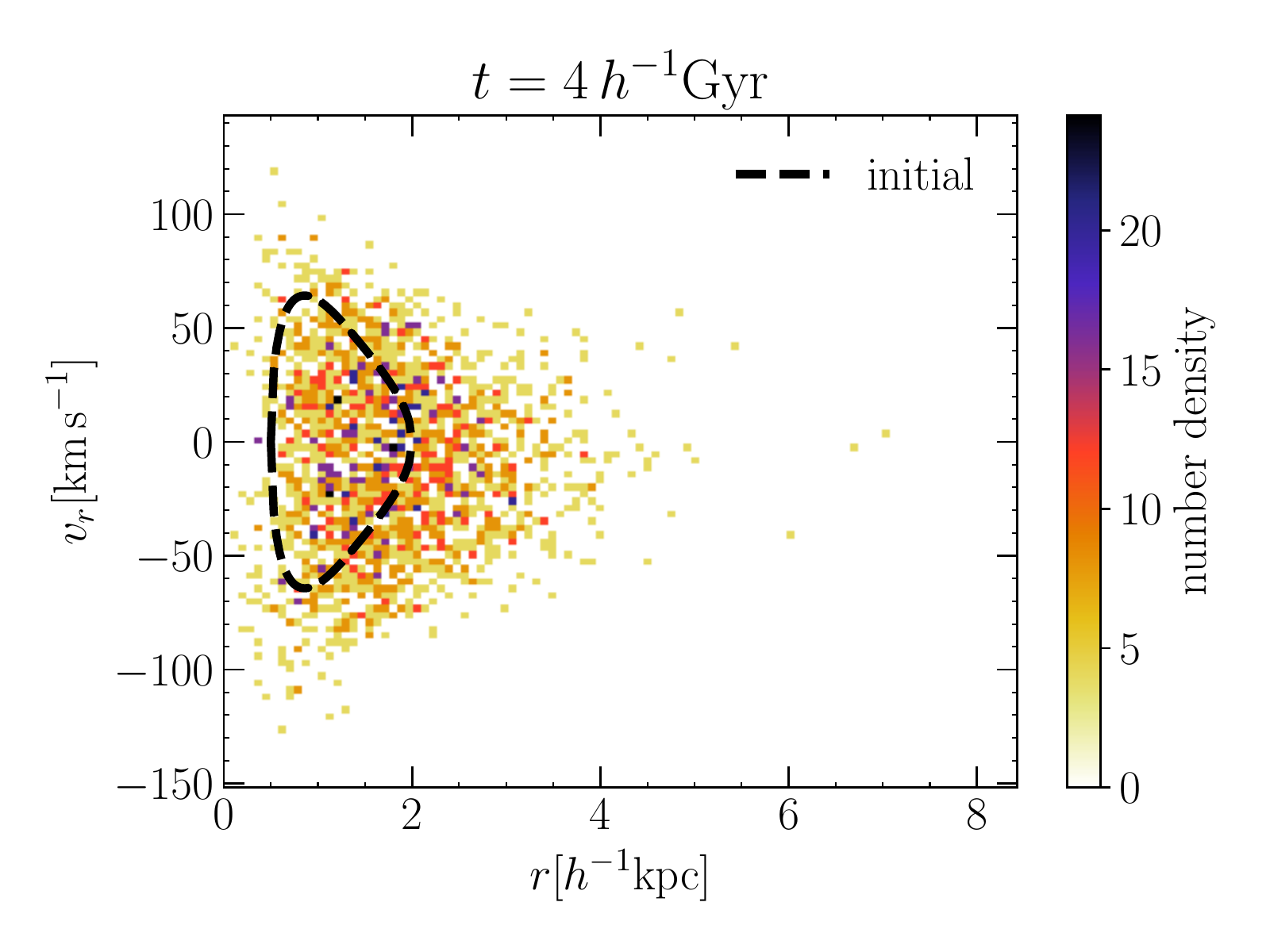}
    \caption{As Figure \ref{fig:Pl_feedback} but for an external disk potential. 
    The upper panels correspond to the simulation in which $\epsilon = 0.01$ (adiabatic), $f_{\rm g} = 0.46$ (adiabatic), and $H = 1.4\,h^{-1}{\rm kpc}$ (adiabatic). The lower panels correspond to the simulation with $\epsilon = 0.05$ (benchmark), $f_{\rm g} = 0.023$ (benchmark), and $H = 0.35\,h^{-1}{\rm kpc}$ (impulsive).}
    \label{fig:disk_feedback}
\end{figure*}

The lower panels of Figure \ref{fig:Pl_feedback} correspond to a simulation in which the Plummer sphere is more compact
(see Table \ref{tab:simu_params}), whereas the injection time and the amount of injected energy assume their benchmark values. As we can see, the results of this simulation are vastly different from those in the upper panel. In the lower left panel 
we see how the halo forms a core of size $\sim 1\,h^{-1}{\rm kpc}$ already during the first explosion cycle, and how it retains this core until the end of the simulation. The final phase space distribution of the orbital family of tracers (lower right panel) 
shows clear signs of an impulsive change in the gravitational potential. The orbital family has split up into several shells and the radial range occupied by the tracers has expanded to significantly larger radii compared to the initial distribution, which is roughly given by the black dashed line. This implies that radial actions are not 
approximately conserved throughout the simulation (as they are in the upper right panel). 

Figure \ref{fig:Pl_feedback} suggests that there is a link between whether or not periodic SN feedback-like energy injection is a feasible core formation mechanism and whether the induced change in the central gravitational potential is adiabatic or impulsive. 

\subsubsection{Feedback from an exponential disk}\label{sec:fbdisk}
\begin{figure*}
    \centering
    \includegraphics[width=0.48\linewidth]{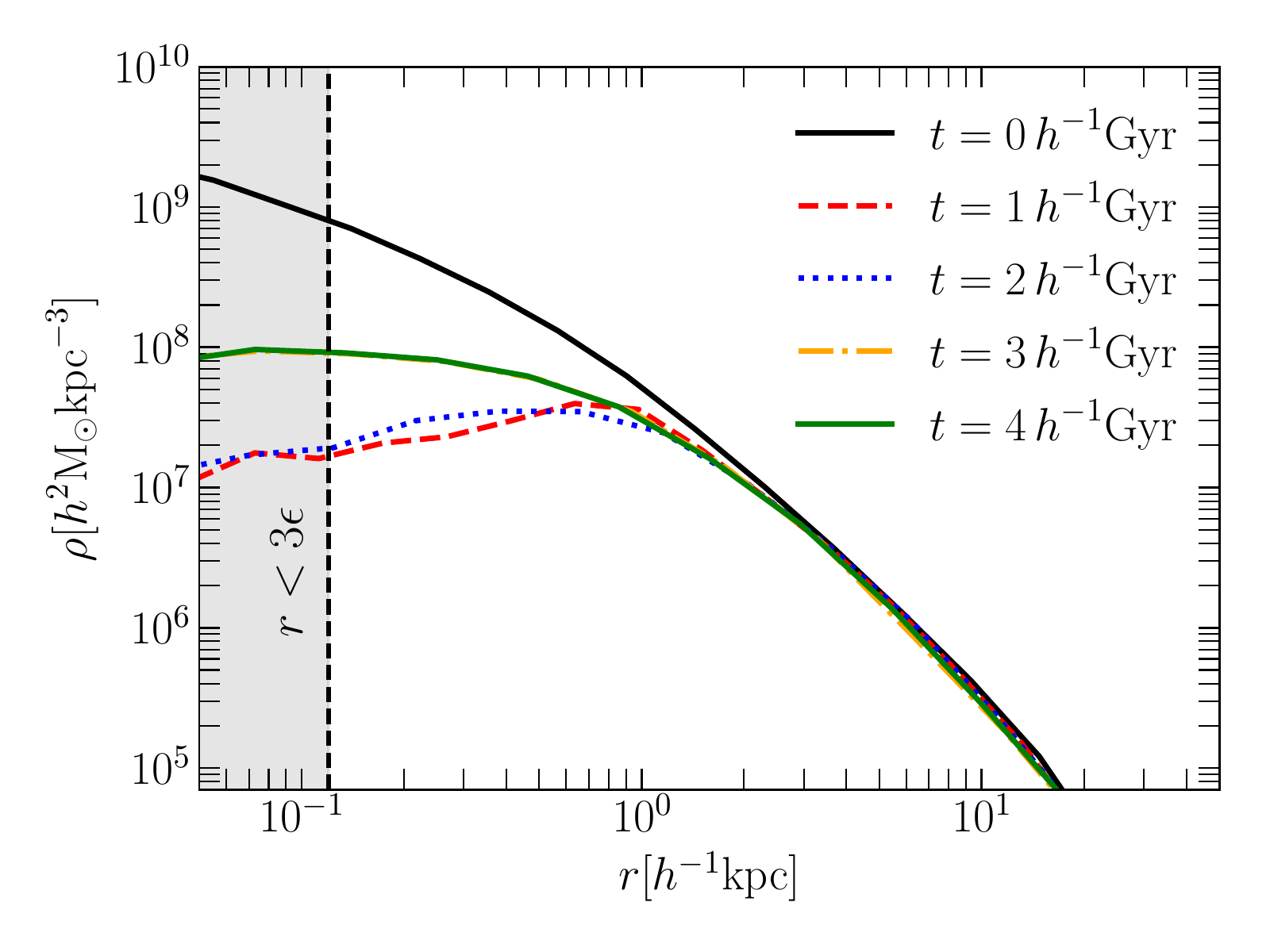}
    \includegraphics[width=0.51\linewidth]{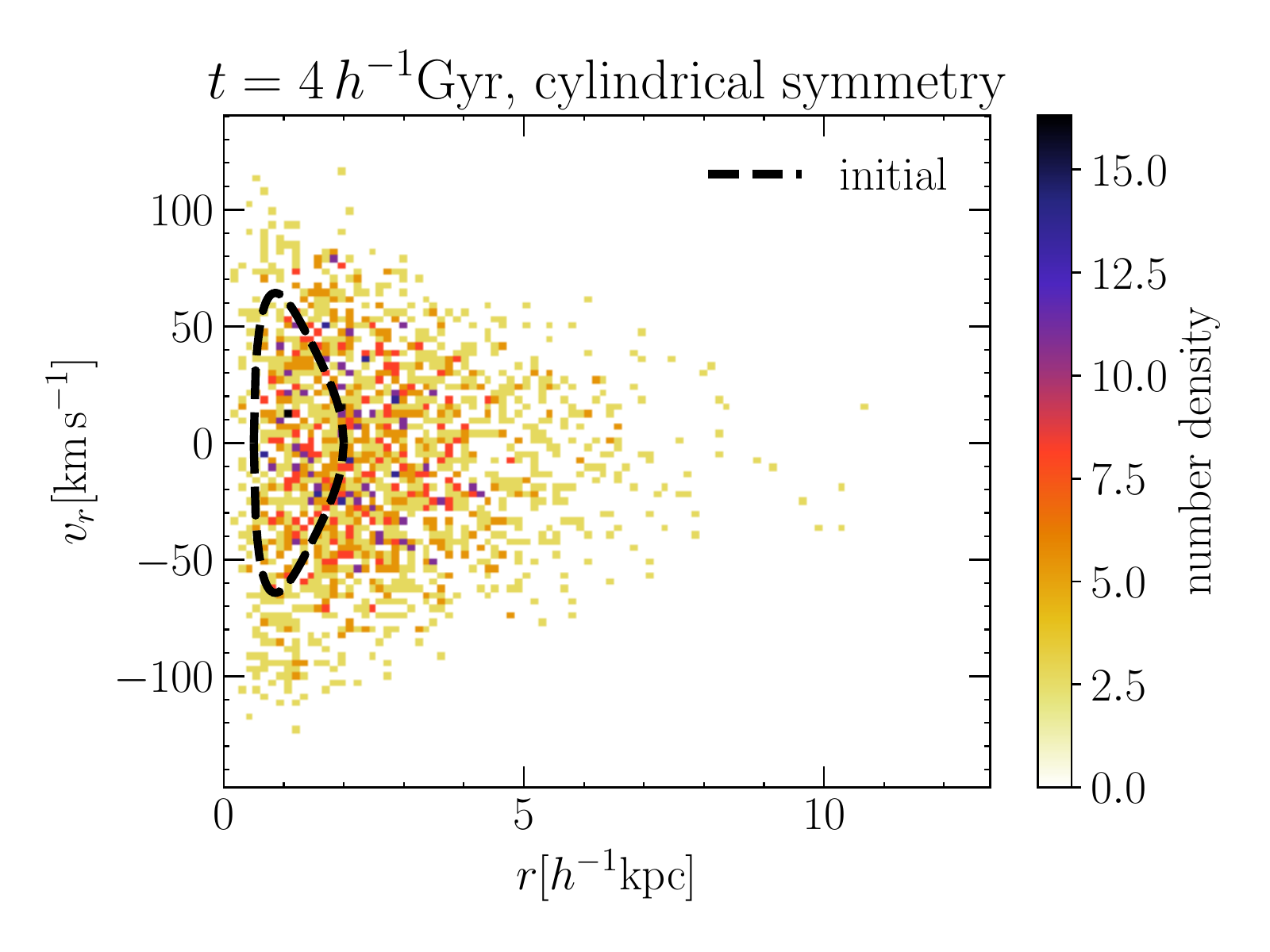}\\
    \includegraphics[width=0.48\linewidth]{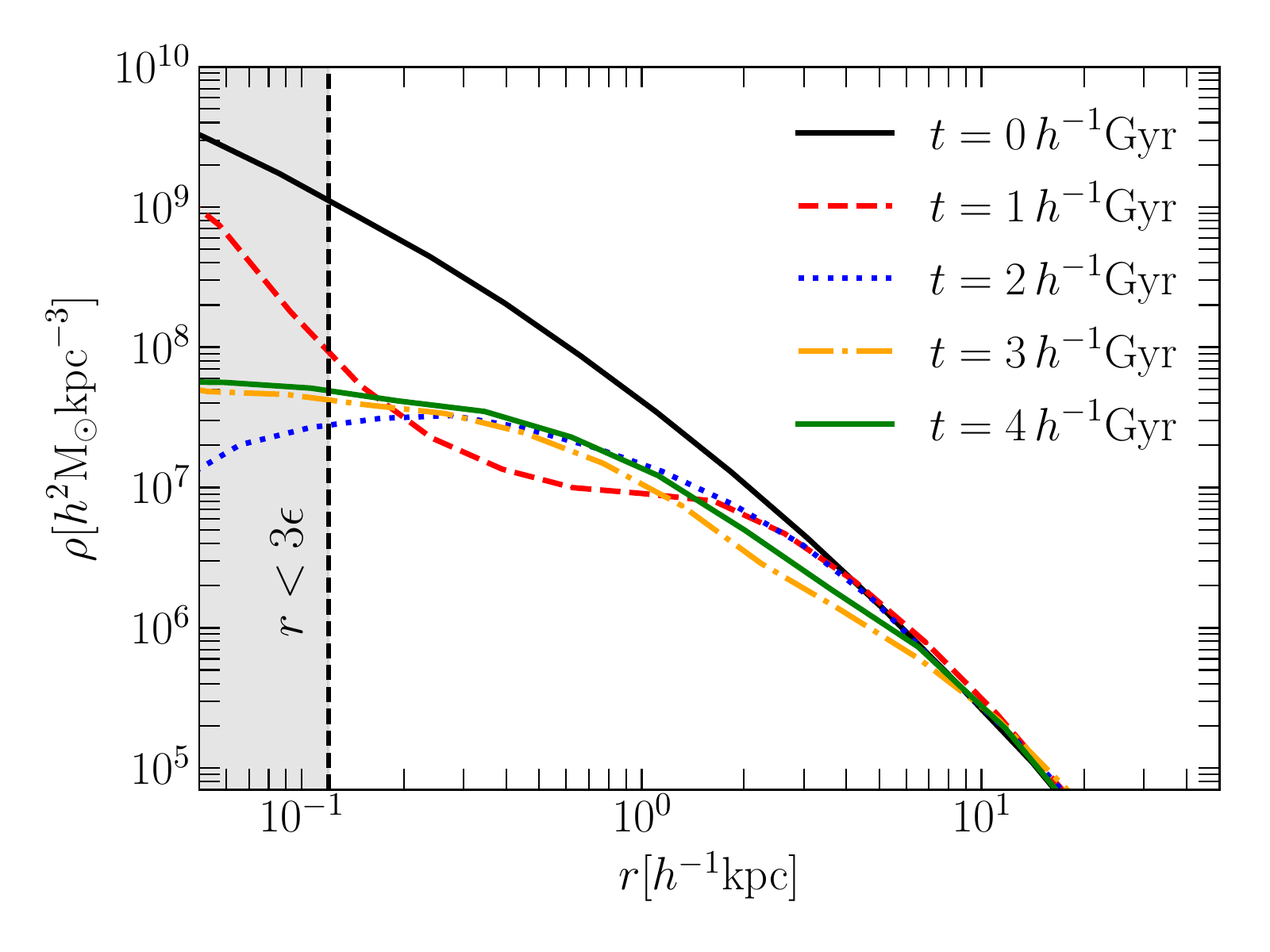}
    \includegraphics[width=0.51\linewidth]{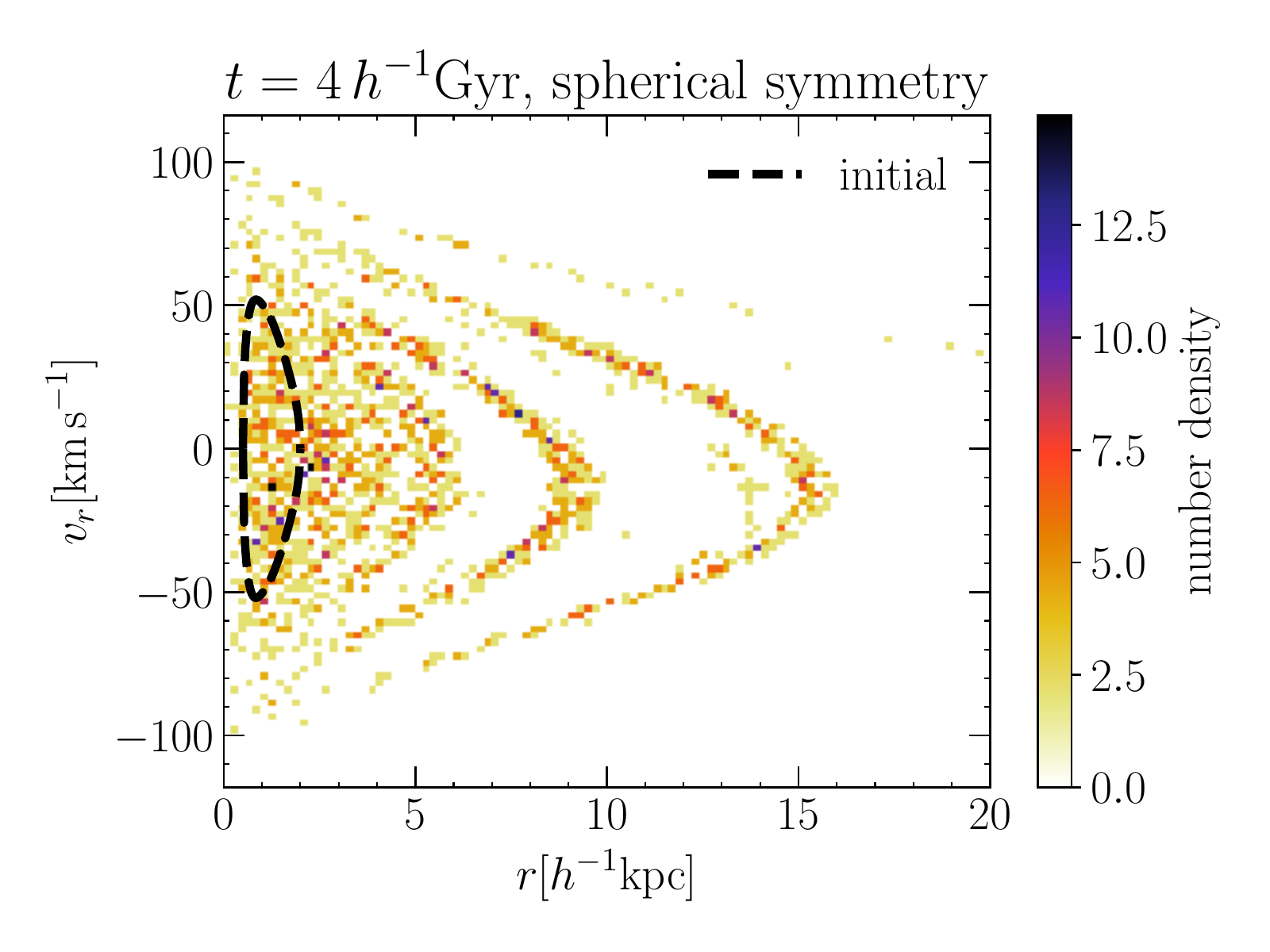}
    \caption{Comparison of the evolution of the DM density profile (left column) and the final phase space distribution of an orbital family of kinematic tracers (right column) between two simulations with external potentials that correspond to identical enclosed mass profiles. The top panels show results from the run including an external disk potential with $\epsilon = 0.4$ (impulsive), $f_g = 0.46$ (adiabatic), and $H = 0.35\,h^{-1}$kpc (impulsive). The bottom panels show results from a simulation in which the ``baryonic'' enclosed mass profile is idential, but generated by a spherically symmetric mass distribution. The distribution of individual ``supernovae'' is sampled from the spherically symmetric ``baryonic'' mass distribution. Notice that while an extended core forms in both cases, the final phase space distribution of the kinematic tracers is distinctly different. } 
    \label{fig:disk_sph_comp}
\end{figure*}

Figure \ref{fig:disk_feedback} shows the same kind of comparison as Figure \ref{fig:Pl_feedback}, but 
for an external disk potential. In the upper panels we 
show the results of the simulation in which all of the relevant parameters in Table \ref{tab:simu_params} correspond to an adiabatic configuration. Just as in the Plummer case, we hardly see any evolution in the DM density profile beyond the resolved radii. 
The final phase space density of the orbital family (upper right panel) does, however, exhibit a degree of orbital diffusion that clearly exceeds the diffusion observed in the upper right panel of Figure \ref{fig:Pl_feedback}. Still, the range of radii covered by the tracers' orbits does not expand much further than the original configuration, which means that there is no net migration outward. 
We attribute the additional diffusion in the orbital family to the 
additional challenges of preserving  cylindrical symmetry.
In fact, the setup of the orbital family relies on a potential which has been measured in the $x-y$ plane. Thus, it should in theory be applicable only to particles whose orbits are confined within the disk plane at all times. However, the slightest perturbation into the vertical direction (notably, our implementation of SN feedback can introduce these) can cause the tracers to obtain a non-zero vertical velocity. The tracers' plane of motion then changes, and the in-plane potential (measured as a function of polar radius) no longer determines the tracers' orbits. This deviation from cylindrical symmetry is thus cause of an additional diffusion of the orbital family. 

The lower panels of Figure \ref{fig:disk_feedback}, shows the case 
in which the disk scale length is a factor of two smaller than the benchmark value 
whereas the injection time and the amount of injected energy take their benchmark values. As in the lower panels of Figure \ref{fig:Pl_feedback}, we can now observe the formation of a $\sim 1\,h^{-1}{\rm kpc}$ core in the DM halo by tracking the evolution of its density profile. 
However, in this case it takes at least two explosion cycles for the core to fully form, indicating that core formation is slightly less efficient for this disk configuration than for the Plummer sphere case.
Moreover, we observe a clear difference between the ``impulsive" simulation that includes the external disk potential and the one including the external Plummer potential when looking at the final phase space distribution of the tracers that were initially part of one orbital family. While the range of radii covered by the tracers expands roughly by the same amount, we do not observe any emergent shell-like patterns in phase space, i.e., a split into several orbital families, in the simulation including an external disk potential. Instead, we find that the final phase space distribution is essentially phase-mixed, indicating that significant diffusion has occurred. As in the adiabatic case, we attribute this to the growing impact of deviations from cylindrical symmetry that accumulate throughout the simulation. We take a closer look at the role of symmetry in Section \ref{subsec:sym}. We thus conclude that while radial migration outwards is a clear signature of impulsive changes to the underlying gravitational potential, shell-like structures, as seen in Figure \ref{fig:Pl_feedback}, are only relatively long-lived (and thus apparent at the end of our simulations) if the potential's underlying spatial symmetry is closely preserved as the potential changes. 

\subsection{The role of symmetry}\label{subsec:sym}

The results of Sections \ref{sec:fbplummer} and \ref{sec:fbdisk} suggest that the symmetry of the external ``baryonic'' potential -- and in turn the distribution of individual ``supernovae'' -- affects the final phase space structure of orbital families. In particular, we find that impulsive SN feedback in simulations with a spherically symmetric external potential gives rise to shell-like features in the phase space of kinematic tracers that initially belonged to the same orbital family; these features are long-lived and remain evident at the end of our simulations (see Figure \ref{fig:Pl_feedback}). However, such long-lasting shell-like features are not present at the end of our impulsive SN feedback simulations with a disk-like external potential (see Figure \ref{fig:disk_feedback}).

In Section \ref{sec:fbdisk} we stated that this divergent behaviour of the kinematic tracers can likely be explained by the difference in spatial symmetry. However, there is also a significant mass difference between the Plummer spheres and the disks. To verify that the decisive factor is symmetry -- and not the mass of the external potential -- we performed an additional set of simulations. For this set, the setup of the external potentials is such that the spherically averaged mass profiles associated with them are identical to the mass profiles in the corresponding runs with external disk potentials, i.e., 
\begin{equation}
    M(r) = M_{\rm d}\left\{1-\left(1+\frac{r}{H}\right)\exp\left(-\frac{r}{H}\right)\right\},
\end{equation}
where $M_{\rm d}$ is the total mass and $H$ is the scale length of the equivalent disk potential. The radii of individual ``supernovae'' are randomly sampled from the normalized mass profile. Since the external potential is now spherically symmetric, the disk height parameter is superfluous. All other parameters are kept as in Table \ref{tab:simu_params}. 

Across simulations, we find that a spherically symmetric external potential induces a qualitatively different contraction of the halos' density profiles. While an axisymmetric disk potential leads to shallower central slopes (see Figure \ref{fig:ICgeneration}), an equivalent spherically symmetric potential gives rise to steeper 
density profiles of all simulated DM halos,
resulting in deeper potential wells, and thus, requiring more energy to unbind the cusps. As a consequence, we find that in such (spherical) cases, no cores form in runs in which $\epsilon = 0.01$ or $\epsilon = 0.05$. Only for the largest choice of the energy coupling parameter, $\epsilon = 0.4$, do cores form. This is the case we choose to 
make the comparison between the axisymmetric and spherical potentials. 
The top row of Figure \ref{fig:disk_sph_comp} shows results from the run with an external disk potential, using the parameters $\epsilon = 0.4$, $f_g = 0.46$, and $H = 0.35\,h^{-1}$kpc. The bottom row shows results of the corresponding simulation including a spherically symmetric external potential and a spherical distribution of ``supernovae''. In the left column, we show the evolution of the spherically averaged DM density profiles, while the final phase space distribution of the orbital family of kinematic tracers is shown in the right column. Although extended constant density cores form in both cases, it is evident that core formation is slower in the spherically symmetric case -- a strong cusp-restoring contraction effect due to the external potential can be observed after $1\,h^{-1}{\rm Gyr}$. Nevertheless, the final core is somewhat larger in the spherically symmetric case. The final phase space distribution of the kinematic tracers is remarkably different between the two simulations. In the 
axysimmetric case, 
we observe a considerable radial expansion on average, in line with the radial expansion of the DM particles. Moreover, the final distribution in radial phase space is largely featureless, i.e. completely phase-mixed. In the spherical case, 
the radial expansion is accompanied by the emergence of prominent shell-like structures. We thus conclude that the divergent behaviour of the kinematic tracers between Figures \ref{fig:Pl_feedback} and \ref{fig:disk_feedback} is due to the difference in spatial symmetry -- and not due to the difference in the baryonic mass.

The reason for this difference is as follows. Initially, all particles are on orbits defined by nearly identical actions ($J_r$ in the case of a spherically symmetric external potential and $J_R$ in the case of a disk), but with different orbital phases. As outlined in \citet{Pontzen:2011ty} and \citet{2019MNRAS.485.1008B}, a sudden change in the gravitational potential causes a change in the energies of kinematic tracers that depends on their respective orbital phases. Thus, impulsive mass removal, as in our effective model of SN feedback, turns the original orbital family into a particle distribution which is not phase-mixed. The shell-like features seen in the bottom right panel of Figure \ref{fig:disk_sph_comp} are signatures of early-stage phase mixing (see e.g. \citealt{2008gady.book.....B}). As long as the underlying symmetry is preserved, i.e, $J_r$ or $J_R$ are integrals of motion for individual tracers, phase mixing progresses relatively slowly. Deviations from spherical (or cylindrical) symmetry, however, can cause orbital diffusion along resonant directions. As a result, actions associated with the broken symmetry are no longer integrals of motion (see \citealt{Pontzen:2015ova}) and phase mixing progresses much faster\footnote{In such cases, these signatures can still be observed in the immediate aftermath of an impulsive energy injection. However, observing them requires a more in-depth analysis (see Burger et. al., in prep.)}. As we mentioned in Section \ref{sec:fbdisk}, the cylindrical symmetry in our runs with an external disk potential is only exact in the $x-y$ plane -- and is easily broken by the distribution of ``supernovae'' during a given explosion cycle. As a result, tracers acquire non-zero vertical velocities and their plane of motion becomes tilted with respect to the disk plane. At that point $J_R$ is no longer an integral of motion and resonant diffusion can occur. Therefore, the divergent behaviour of the tracers in Figures \ref{fig:Pl_feedback}, \ref{fig:disk_feedback}, and \ref{fig:disk_sph_comp} is due to the difference in both the spatial symmetry of the external potential and the spatial distribution of individual ``supernovae''.  


\subsection{Net SN feedback impact on the inner DM distribution} \label{sec:netsnfimp}
\begin{figure*}
    \centering
    \includegraphics[width=0.49\linewidth]{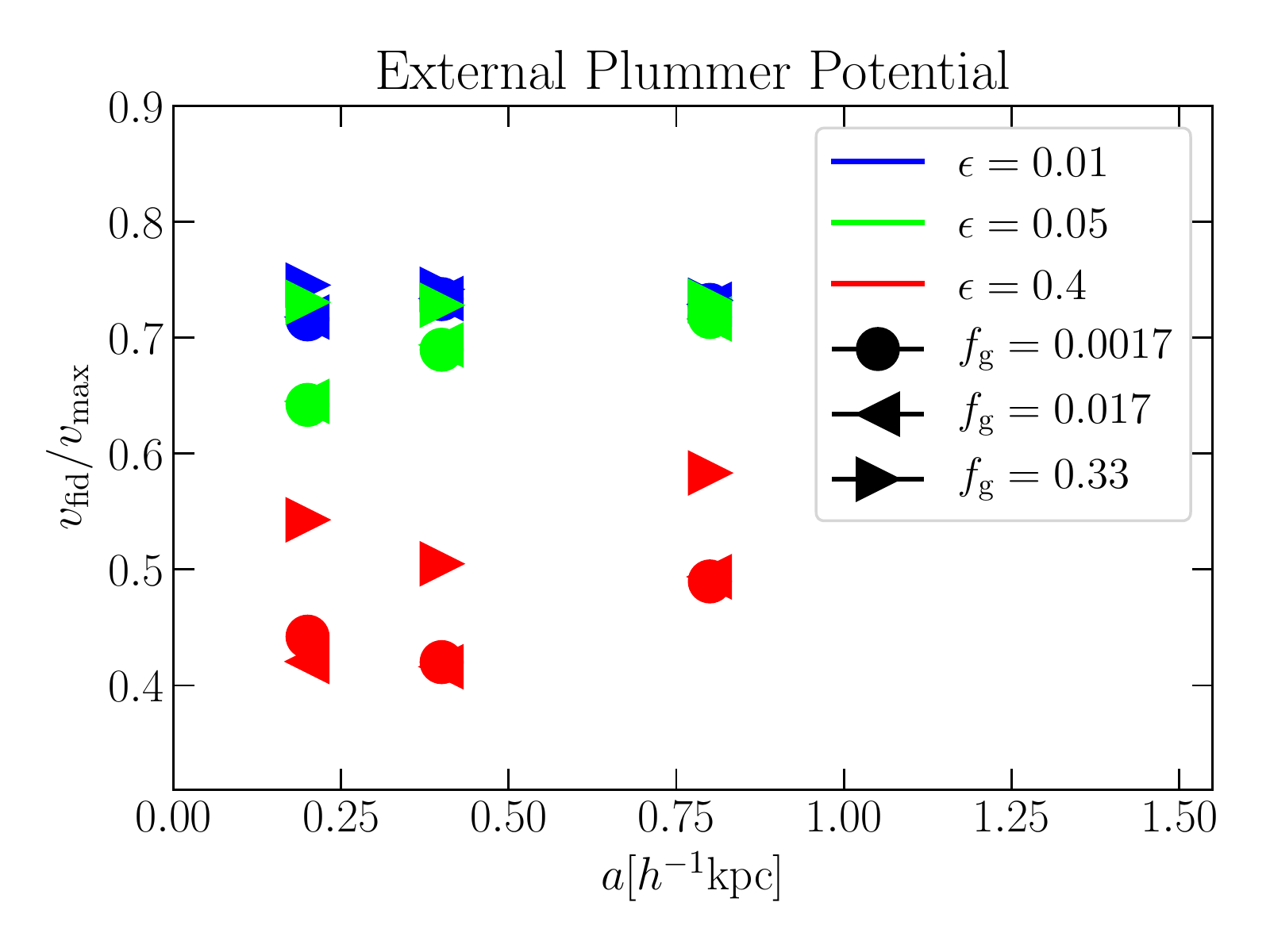}
    \includegraphics[width=0.49\linewidth]{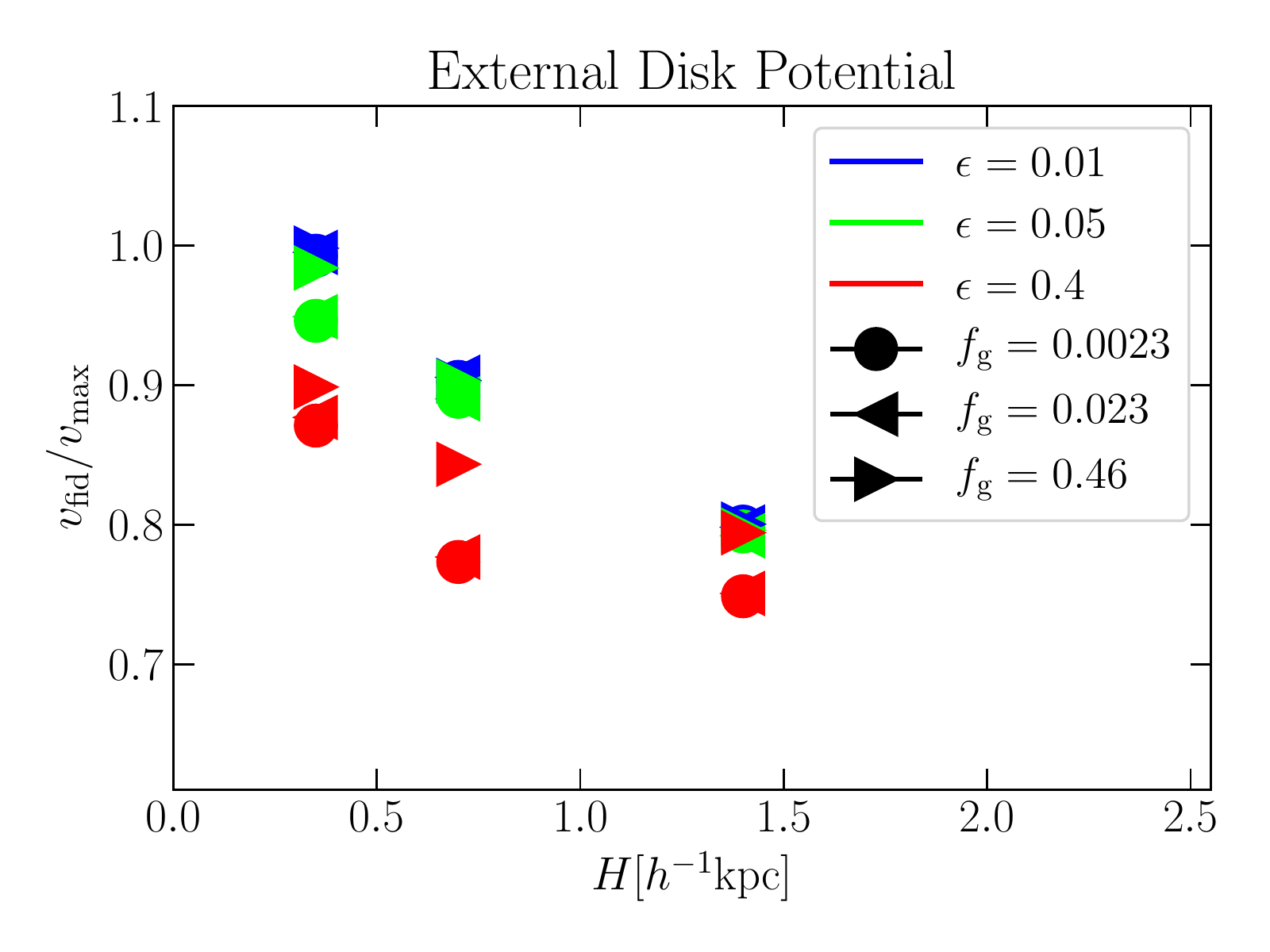}
    \caption{A comparison showing the impact of SN feedback on the inner DM distribution in all 27 simulations including an external Plummer and disk potential on the left and right panels, respectively. The $y-$axis quantifies how cuspy/cored the total matter density profile is, i.e., how fast the galaxy circular velocity curve rises, depending on the SN feedback configuration. Lower values of $v_{\rm fid}/v_{\rm max}$ correspond to more cored profiles. However, this ratio can also be altered by contraction of the halo due to the external potential (prior to the first SN feedback cycle; see \ref{fig:ICgeneration}), an effect that is of particular importance for the SMC-like galaxy (right panel). For that reason, we plot the scale length of the external potentials on the $x-$axis. At fixed values of the scale length, we can then 
    make a relative comparison between the different SN feedback configurations.
    Besides galaxy size, we also show how $v_{\rm fid}/v_{\rm max}$ depends on total injected energy and energy injection time, using color-coded symbols as explained in the legends. Across both panels, we find that injected energy and injection time together largely determine whether the DM halo's density profile changes or not. Still, a residual impact of the galaxy's concentration can also be observed.}
    \label{fig:core_size_results}
\end{figure*}

Figure \ref{fig:core_size_results} shows a comparison of the results of all the simulations described in Table \ref{tab:simu_params}. Following \citeauthor{2020MNRAS.495...58S} (2020, see also \citealt{Oman2016}), we classify halo density profiles with a single number by their galaxy's circular velocity curves through the ratio $v_{\rm fid}/v_{\rm max}$ where $v_{\rm fid}$ is the circular velocity at a radius $r_{\rm fid} = 2(v_{\rm max}/70\,{\rm km\,s^{-1})kpc}$ and $v_{\rm max}$ is the maximal circular velocity. As demonstrated in \citet{2020MNRAS.495...58S}, smaller ratios between the two velocities correspond to more cored DM density profiles. The value for a NFW profile is typically $v_{\rm fid}/v_{\rm max}\sim 0.7$. We note however that there is no 
absolute correspondence between the ratio $v_{\rm fid}/v_{\rm max}$ and how cored the DM halo's density profile is 
since this ratio depends on halo concentration (for a fixed $v_{\rm max}$), and the precise shape of the profile (e.g. a Hernquist profile, albeit cuspy, has
slightly different values of $v_{\rm fid}/v_{\rm max}$). 
More importantly, the baryonic galaxy can have a major impact on this value in several ways: i) massive and concentrated galaxies contribute significantly to the total circular velocity curve in the inner region and can contract the DM halo, making it cuspier in the center (and thus increasing its contribution to the circular velocity curve); ii) SN-driven outflows expel gas from the center and redistribute the DM from the inside out, reducing the value of $v_{\rm fid}/v_{\rm max}$, which is the focus of this work.
In our simulations, the mass of the external potential is the same across all simulations of a given type (Plummer or disk) and thus, since the scale length of the galactic potential appears on the $x-$axis of Figure \ref{fig:core_size_results}, all measured ratios $v_{\rm fid}/v_{\rm max}$ with the same $x-$coordinate correspond to simulations with the same initial conditions. Hence, while the value of $v_{\rm fid}/v_{\rm max}$ cannot tell us in absolute terms which DM halo is more cored {\it across all simulations}, we can use
the difference between measured values of $v_{\rm fid}/v_{\rm max}$ with the same $x-$coordinate to establish which combination of parameters in our SN feedback model (see Table \ref{tab:simu_params}) is more efficient at forming a core.   

On the left panel of Figure \ref{fig:core_size_results}, we show the results of the 27 simulations including an external Plummer sphere. As can be seen in Figure \ref{fig:ICgeneration}, the mass of the Plummer sphere chosen to mimic a Fornax-like dwarf galaxy is not large enough to cause a significant contraction of the inner DM halo, allowing for an easy comparison of the circular velocity curves between simulations.
We can see that the measured final values of $v_{\rm fid}/v_{\rm max}$ are very similar for all choices of $a$, as long as the injected energy is small and the injection time is long, i.e., small $\epsilon$ and large $f_{\rm g}$ (see Table \ref{tab:simu_params}). Thus, for this combination of $\epsilon$ and $f_{\rm g}$ (both assuming their ``adiabatic'' values) the halo essentially retains its cusp for all choices of the galaxy's concentration. 
The most important factor in determining whether or not a core is formed is the amount of injected energy. In fact, if the a priori energy coupling parameter is small ($\epsilon=0.01$), a (very small) core is formed only if the galaxy is very concentrated and the injection time is quite short. However, if the energy coupling is very large ($\epsilon=0.4$), cores are formed for virtually every combination of the other two parameters. Injection time and concentration of the galaxy play somewhat smaller roles, with injection time appearing to be slightly more important: the core becomes more significant the shorter the injection time is and/or the more concentrated the Plummer sphere is.  
We note that varying the injection time between 1 per cent and 10 per cent of the orbital family's radial period ($f_g=0.0017$ and $f_g=0.017$, respectively) produces virtually no difference. 
This indicates that for an intermediate value of the energy coupling parameter, the relevant timescale for energy injection is roughly set by the orbital period of particles in the halo's center (and not much smaller than that), suggesting that although SN feedback needs to be impulsive for a core to form, the requirement for the degree of impulsiveness (and thus for how bursty star formation should be) is not that severe. 

The right panel of Figure \ref{fig:core_size_results} shows the case 
of the external disk potential introduced to mimic a SMC-like galaxy. From Figure \ref{fig:ICgeneration} we know that this far more massive external potential causes a significant contraction of the DM halo, which is reflected in the values of $v_{\rm fid}/v_{\rm max}$ (compared to left panel of Figure \ref{fig:core_size_results}).
In fact, all of the measured values lie above the value 0.7, the value for a typical unperturbed NFW halo. However, it is obvious from Figure \ref{fig:disk_feedback} that cores are formed in our simulations with a disk as long as SN feedback is sufficiently energetic and impulsive. As we discussed at the beginning of this subsection, the absolute values of $v_{\rm fid}/v_{\rm max}$ do not imply on their own whether a halo is cored or not.
We have explicitly verified that no cores form in the runs in which $\epsilon = 0.01$ and $f_{\rm g} = 0.46$. This implies that for a fixed value of $H$, we can take the blue right-pointing triangle as a 'cuspy' baseline and assess how cored the DM profile is for the other simulations at the
same $H$ (but with different parameters regulating injected energy and injection time). 
With that in mind, we observe very similar trends as the ones for the Plummer sphere on the left panel of Figure \ref{fig:core_size_results}.
In general, however, the final profiles are somewhat less cored on average for the SMC-like case than for the Fornax-like case, relative to the baselines (blue right-pointing triangle). 
If cores form for a given combination of $\epsilon$ and $H$, the core size (significance) is again regulated by how short the injection time is. The same is true for the disk's scale length $H$, with more significant cusp-core transformation (relative to the baseline) for more concentrated disks. 

\subsection{Actual and nominal energy change}\label{sec:reseng}
\begin{figure*}
    \centering
    \includegraphics[width=0.49\linewidth]{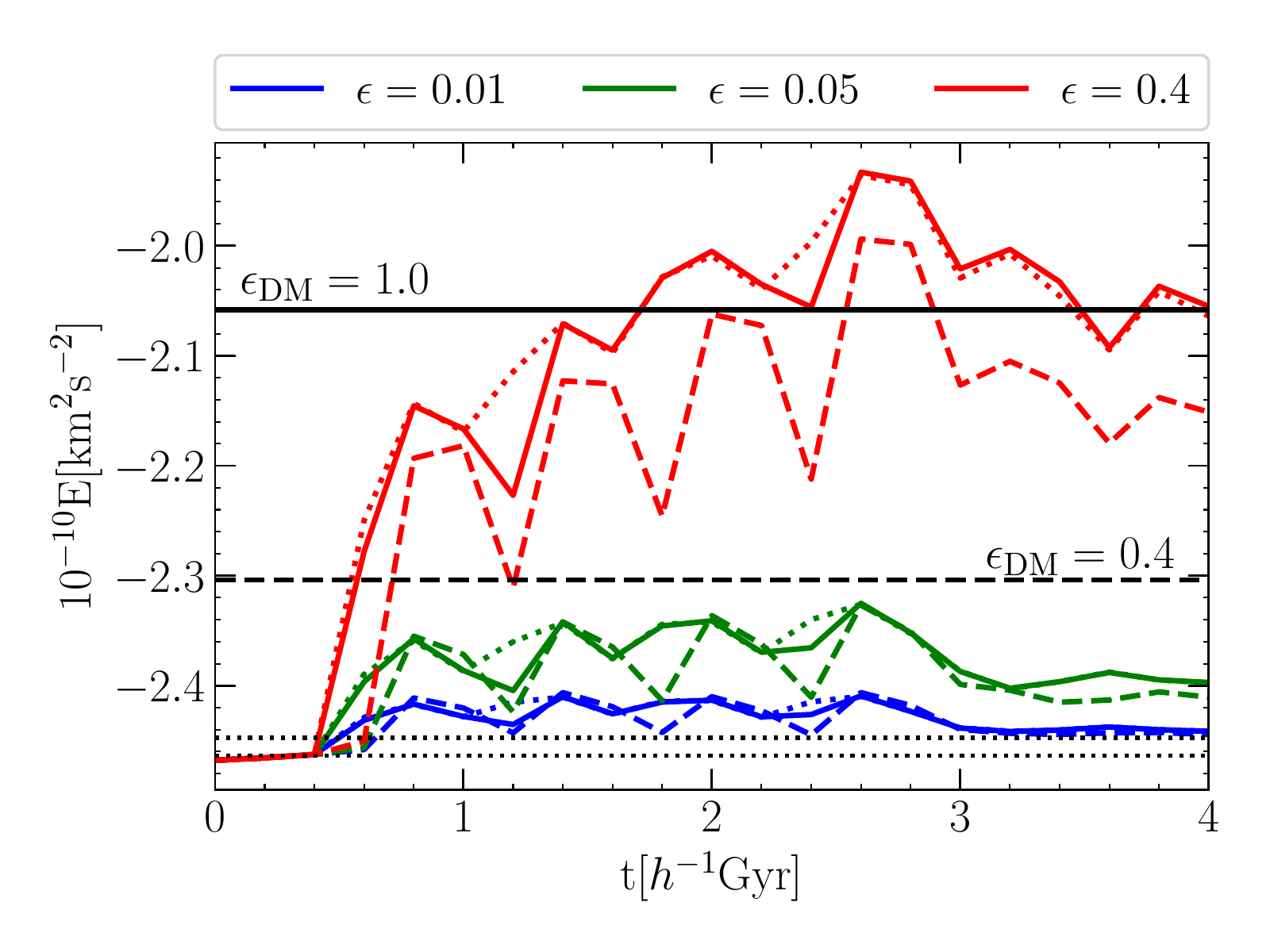}
    \includegraphics[width=0.49\linewidth]{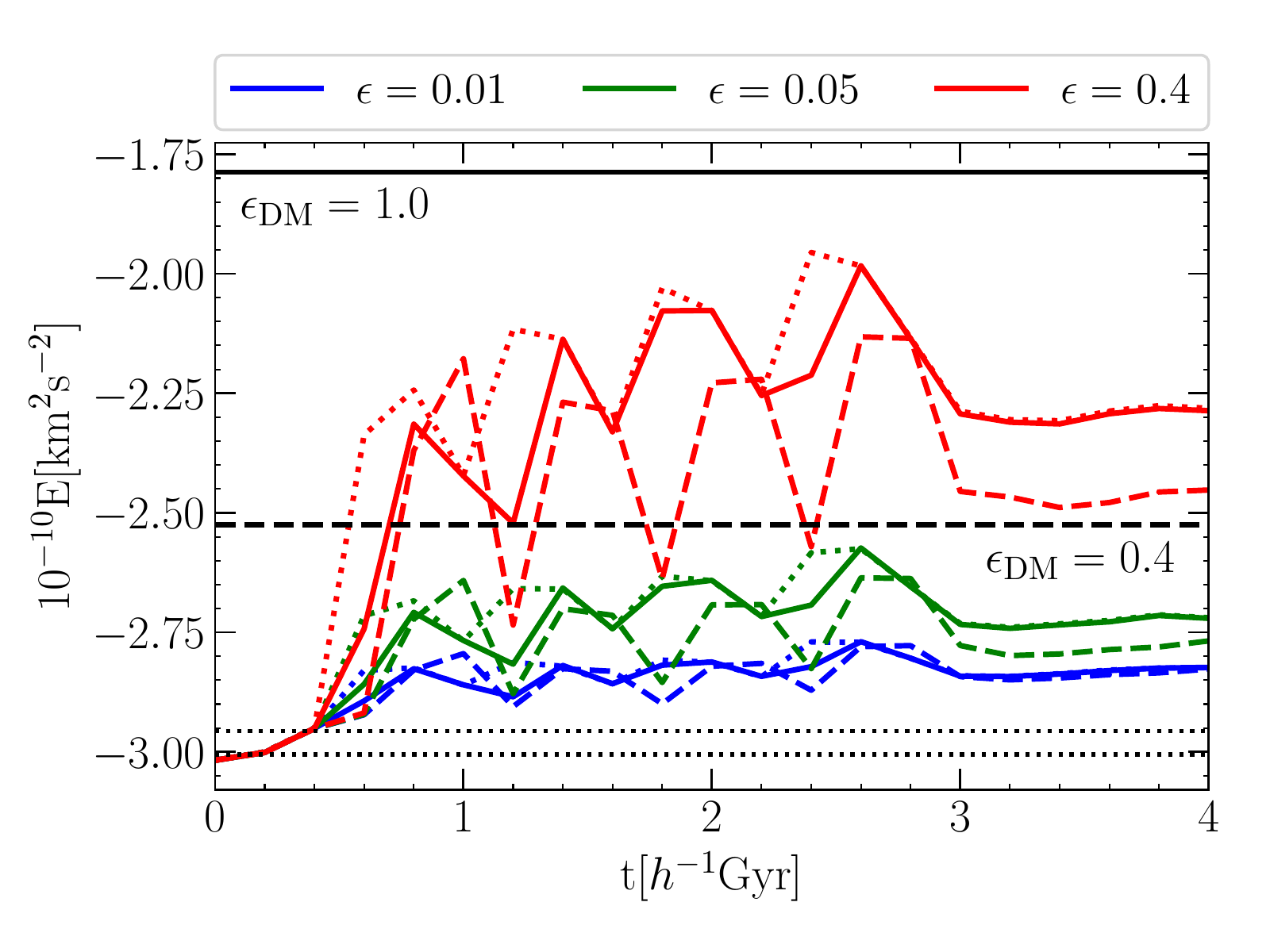}
    \caption{The total energy (per unit mass) of DM partices in the halo as a function of time in the nine simulations including an external Plummer sphere with $a = 0.4\,h^{-1}$kpc (left panel) and an external exponential disk with $H = 0.35\,h^{-1}$kpc (right panel). The colours of the lines refer to simulations with nominal energy couplings $\epsilon$ as indicated in the legends above the panels. Solid coloured lines refer to the runs using the benchmark value of $f_g$, dashed (dotted) coloured lines to the runs using the adiabatic (impulsive) values of these parameters (see Table \ref{tab:model_params}). Black horizontal lines show the final energies corresponding to ``effective'' energy couplings ($\epsilon_{\rm DM}$, see Equation \ref{eq:edm}) of 1.0, 0.4, 0.05, and 0.01, respectively. The ``effective'' energy coupling is larger than the nominal coupling in all simulations, suggesting that our order of magnitude estimate for the injected energy systematically underestimates the actually injected energy. Note that the summed energy of the DM particles is approximately stable during the final $1\,h^{-1}$Gyr for almost all shown simulations, with the exception of the three runs with the largest nominal energy coupling and an external Plummer potential (red lines on the left panel). }
    \label{fig:ecomp}
\end{figure*}

As shown in Section \ref{sec:netsnfimp}, the sizes of the final DM cores vary for a fixed value of the nominal energy coupling parameter $\epsilon$. Assuming that the final DM halos are in dynamical equilibrium and fulfill the virial theorem, this implies that the total energy that has been transferred to the DM particles in the halo is different between those simulations (\citealt{Penarrubia:2012bb}). The effective coupling between SN feedback and DM, $\epsilon_{\rm DM}$, is thus distinctly different from the nominal energy coupling $\epsilon$ defined in Section \ref{sec:sn_imp}. 

In Figure \ref{fig:ecomp} we show the energy that is actually injected into the DM halo for the nine simulations with an external Plummer profile and $a = 0.4\,h^{-1}$kpc on the left panel and the nine simulations with an external disk potential and $H = 0.35\,h^{-1}$kpc on the right panel. The energies displayed correspond to the total energy of all DM particles in the halo at a given time measured from snapshots taken every $200\,h^{-1}$Myr. 
For orientation, we also show as horizontal lines final energies corresponding to several values of the effective coupling parameter $\epsilon_{\rm DM}$, which we define as
\begin{equation}
    \epsilon_{\rm DM} = \frac{\langle m_\star\rangle\Delta E_{\rm DM}}{M_\star \xi(m_\star > 8M_\sun)E_{\rm SN}}, \label{eq:edm}
\end{equation}
where $\Delta E_{\rm DM}$ is the increase in the DM halo's total energy and all other quantities are as in Equation \ref{eq:penarrubia}. The values shown for $\epsilon_{\rm DM}$ are $1.0$ (solid black line), $0.4$ (dashed black line), $0.05$, and $0.01$ (thin dotted black lines). 
For individual runs, we can compare the nominal energy coupling (to the ISM) $\epsilon$ with the actual, effective energy coupling $\epsilon_{\rm DM}$ at the end of the simulation.

In the Plummer case (left panel of Figure  \ref{fig:ecomp}), we see that our effective model underestimates the energy that is truly injected into the halo in all nine simulations. The discrepancy between $\epsilon$ and $\epsilon_{\rm DM}$ is particularly large for small nominal energy couplings and decreases for larger values of $\epsilon$. This is in line with our remarks in Section \ref{sec:sn_imp}. The approximation that the injected energy is given by the binding energy associated with one of the Plummer spheres performs better if the local density contrast generated by individual ``supernovae'' is large.  

This picture is validated in the disk case (right panel of Figure \ref{fig:ecomp}), where the agreement between nominal and actual injected energy is again better in the three simulations in which the nominal energy coupling $\epsilon = 0.4$. We can explain this behaviour by recapitulating how we implement the energy injection in our effective model of SN feedback. To fix the mass of an individual ``superbubble'', we assume that the energy that is required to unbind a Plummer sphere of said mass is given by Equation \ref{eq:sn_selfen}, i.e., the gravitational binding energy of such a Plummer sphere in vacuum. However, within the gravitational potential of the halo -- and the external disk or Plummer potential -- removing a Plummer sphere of mass $m_{\rm max}$ is equivalent to an energy injection that is larger than just the binding energy given in Equation \ref{eq:sn_selfen}. This is because the gravitational pull of the surrounding matter needs to be overcome as well. The relative contribution of this extra energy injection, which is associated with the interaction with the surrounding matter and unaccounted for in our model, is larger if the local density contrast generated by the ``superbubble'' is small.  As a consequence, the ratio between $\epsilon_{\rm DM}$ and $\epsilon$ is closer to unity in simulations with a larger nominal energy coupling, as can be seen on both panels of Figure \ref{fig:ecomp}. In line with these considerations, similar trends emerge when comparing between simulations with external Plummer (disk) potentials of different sizes. The effective energy coupling is smaller in simulations with more extended potentials since the ambient density in the surroundings of the ``superbubbles'' is smaller.

A few further general statements can be made from Figure \ref{fig:ecomp}:
\begin{itemize}
\item The order of magnitude estimate of the total injected energy is roughly consistent with the nominally injected energy with the maximum discrepancy being a factor of a few in the Plummer case.
\item In the disk case, the comparison between $\epsilon$ and $\epsilon_{\rm DM}$ is complicated by some residual evolution in the total energy at the beginning of the simulation -- particularly for low nominal energy couplings.
\item The qualitative scaling of the injected energy with the nominal energy injection parameter is as intended, i.e., there are sizeable differences between the three cases. 
\item In almost all simulations, the energy is relatively stable after $3\,h^{-1}$Gyr, i.e., when the last SN cycle ends. An exception are the Plummer runs with $\epsilon = 0.4$. The residual fluctuation here is likely a numerical effect. Violent behaviour of particles in the halo's center can impede the accurate determination of the position and the velocity of the halo's center of potential. 
\item Gravitational coupling of energy to the DM particles is not perfect. This is immediately evident from the fact that for $\epsilon = 0.05$ and $\epsilon = 0.4$, the increase in energy of the DM particles is smaller if the energy is injected over a longer time. 
\end{itemize}
Related to the last point, we notice that DM halos whose final density profiles are more cored (see Figure \ref{fig:core_size_results}) also have larger final energies (for the same external potential). Given that the energy is -- in most cases -- stable after the last explosion cycle, we can assume that the DM halos are once more in virial equilibrium at the end of the simulation. The correlation between injected energy and final core size is therefore expected (see \citealt{Penarrubia:2012bb}).

In summary, our formalism provides a reasonable (within a factor of a few) estimate for the mass that needs to be removed by individual ``supernovae'' to match the nominally injected energy (Equation \ref{eq:penarrubia}). In most cases, the actual injected energy is somewhat larger. The model can safely be used to test the impact of SN feedback of different strengths. However, if knowledge of the exact amount of injected energy is required, then the sum over the energies of the DM particles at the end of the simulation will have to be manually compared to the equivalent sum at the beginning of the simulation.

\subsection{Core formation and impulsive energy injection}\label{sec:core_imp}

We have seen so far that core formation
always coincides with at least one signature of an impulsive change of the gravitational potential in the phase space plots of the orbital family (radial expansion or formation of shells). We should note that while observing these signatures is a necessary condition for core formation, it is not sufficient.
For instance, we observe shell-like structures in the Plummer sphere simulations with small energy coupling if the energy injection time is short and $a < 0.8\,h^{-1}{\rm kpc}$, but this system does not develop a significant core. This implies that the most important criterion for an effective cusp-core transformation is whether or not the total injected energy is sufficient, which is a
requirement calculated in \citet{Penarrubia:2012bb}. If there is enough SN feedback energy deposition, then the significance of the core formed depends on how impulsively the energy injection proceeds (see Figure \ref{fig:core_size_results}). 
In light of that, perhaps the most striking result of Figure \ref{fig:core_size_results} is that if the injected energy is large enough, cores can form even if the injection time is of the order of a typical radial period in the halo's central region (red right-pointing triangles in Figure \ref{fig:core_size_results}). If impulsive energy injection is, as stated above, a prerequisite for core formation, then this implies that the rate at which an injection of energy changes the gravitational potential must be fast enough to be perceived as impulsive by a sizeable amount of particles in the halo's center. Below, we aim to provide a more detailed discussion of this in light of our results in Figure \ref{fig:core_size_results}.

\citet{2020arXiv201200737B} presented a theory for the diffusion of radial actions in time-dependent spherical potentials. One key result is that radial actions $J_r$ in slowly evolving potentials can be written to first order as 
\begin{equation}
    J_r = J_{r'}+(\mathbf{r}\cdot\mathbf{v})\frac{\dot{R}}{R}\frac{P_r(E,L)}{2\pi}\equiv J_{r'}+\Delta J_r,
\end{equation}
where $J_{r'}$ is a dynamical invariant (the radial action in a frame in which the potential is time-independent), $P_r(E,L)$ is a particle's orbital period and $\dot{R}/R$ describes the rate at which the gravitational potential changes. Another important result of \citet{2020arXiv201200737B} is that whether radial action distributions display adiabatic or impulsive evolution can be roughly determined by the ratio $\sqrt{\tilde{D}}/J$, where $\tilde{D}$ is the diffusion coefficient of the distribution of radial actions, which for phase-mixed particle ensembles can be calculated as 
\begin{equation}
    \tilde{D}(J_r,E,t) \approx \frac{1}{2}\left(\frac{\dot{R}}{R}\right)^2\left(\frac{P_r(E,L)}{2\pi}\right)^2\left<(\mathbf{r}\cdot\mathbf{v})^2\right>.\label{eq:bm_diff}
\end{equation}
where the brackets represent an ensemble average. The radial action distributions of ensembles of tracer particles (such as the orbital families here) evolve adiabatically if $\sqrt{\tilde{D}}/J_r \le 0.1$, but their evolution becomes increasingly non-linear as this ratio rises. 

\begin{figure}
    \centering
    \includegraphics[width=\linewidth]{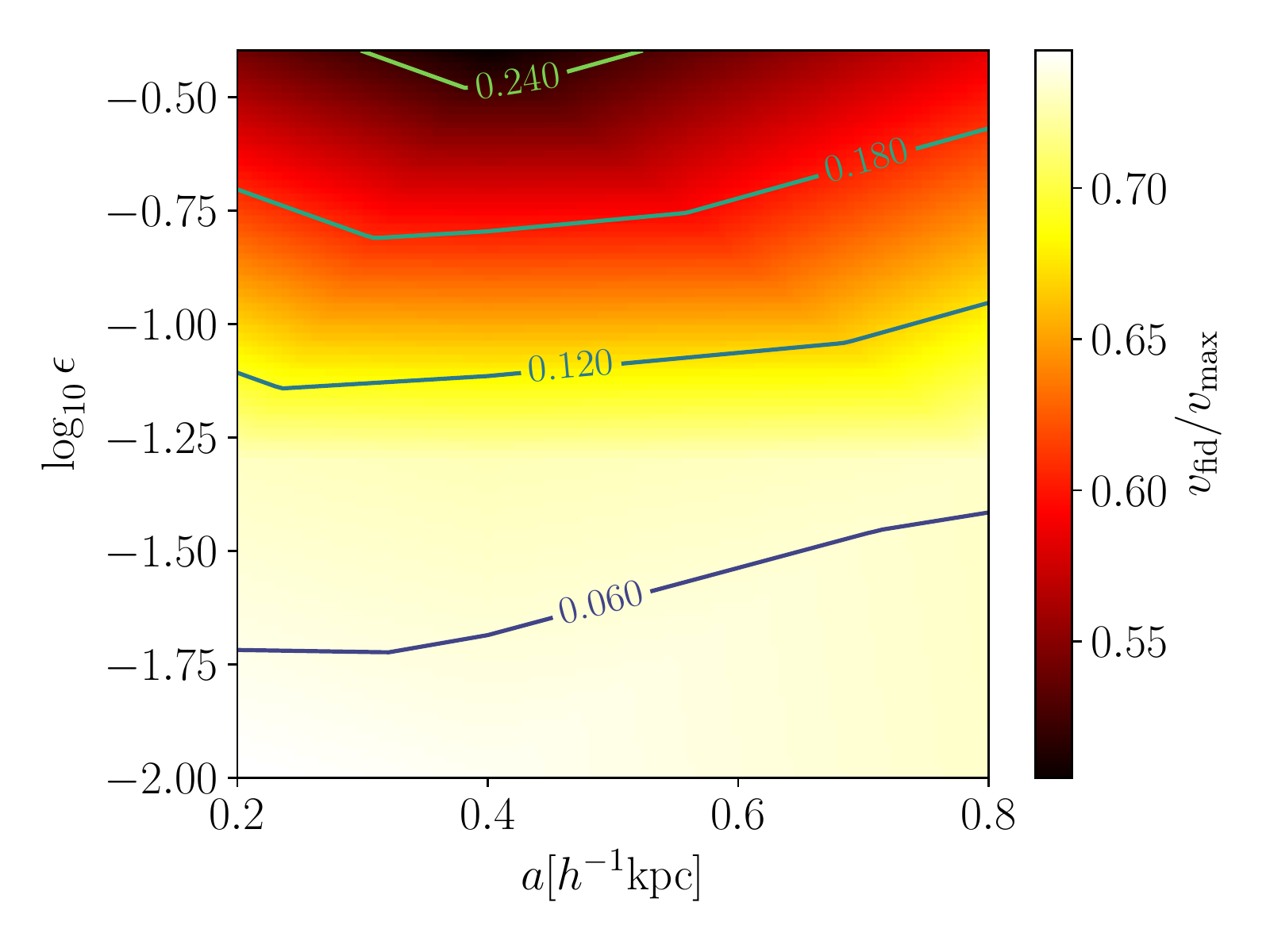}
    \caption{Bilinear interpolation of $v_{\rm fid}/v_{\rm max}$ (a measure of how cuspy/cored the total density profile is; see Figure \ref{fig:core_size_results}) as a function of the scale of the external Plummer potential, $a$, and the logarithm of the energy coupling, $\log_{10}\epsilon_{\rm DM}$. 
    The interpolation is done using the nine simulations including an external Plummer profile and with $f_{\rm g} = 0.33$ (see Table \ref{tab:simu_params}). 
    The lines correspond to a theoretical measure of how impulsive SN feedback is as seen by the tracers comprising the orbital family. It measures the amplitude at which the radial action of a typical tracer oscillates, normalized by the action itself. Larger values correspond to particles whose actions evolve more impulsively on average.}
    \label{fig:overlay}
\end{figure}

In Figure \ref{fig:overlay}, we aim to link this measure of whether or not radial actions evolve impulsively on average to the observed final DM density profiles in the simulations with an external Plummer potential and $f_g = 0.33$ (corresponing to the nominally adiabatic energy injection; see Table \ref{tab:simu_params}). We show a bilinear interpolation of $v_{\rm fid}/v_{\rm max}$ as a function of $a$ and the logarithm of the energy coupling parameter, $\epsilon$, using the nine simulations corresponding to all the right-pointing triangles in the left panel of Figure \ref{fig:core_size_results}. As we have shown earlier, the core becomes more significant with larger energy coupling. To test how this relates to our above-introduced measure of how impulsive the energy injection is, we calculate an estimate of $\sqrt{\tilde{D}}/J_r$ for each of those nine simulations. To do that, we evaluate $J_r$, $P_r(E,L)$ and $\left<(\mathbf{r}\cdot\mathbf{v})^2\right>$ from Equation \ref{eq:bm_diff} for particles which are part of the initial orbital family of tracers. 
To obtain a rough estimate of $\dot{R}/R$, we make use of Equations C6 and C13 in \citet{2020arXiv201200737B}, but assume here that the change in the amplitude of the potential dominates over the change in its logarithmic slope\footnote{We note that this might not be a good approximation since the central slope of the gravitational potential 
can change  substantially during the energy injection. However, taking this effect into consideration
would require an in-depth analysis of individual DM particle's orbits, which is beyond the scope of this paper. Hence, the calculation of $\dot{R}/R$ presented here can be taken as a rough estimate.}. We estimate $\dot{R}/R$ at the pericenter radius of the orbital family, since the change in gravitational potential is more disruptive in the halo's center:
\begin{equation}
    \frac{\dot{R}}{R}\approx \frac{1}{2+\alpha(\,r_0,t_0)}\frac{\Psi(r_0,t_0+f_{\rm g}P)-\Psi(r_0,t_0)}{f_{\rm g}P\Psi(r_0,t_0)},\label{eq:change}
\end{equation}
where $r_0 = 0.5\,h^{-1}{\rm kpc}$, $t_0$ is the time at which the first SN cycle starts, $P$ is the explosion cycle period (see Table \ref{tab:model_params}), and $\alpha$ denotes the logarithmic slope of the shifted potential $\Psi$, which is defined as $\Psi(r,t) = \Phi(r,t)-\Phi(0,0)$. 

Using Equations \ref{eq:bm_diff} and \ref{eq:change}, we can estimate the typical diffusion coefficient $\tilde{D}$ for the orbital family of tracers in each of the simulations with $f_{\rm g} = 0.33$ (slow energy injection). From a bilinear interpolation of the values $\sqrt{\tilde{D}}/J_r$ obtained from each of those runs we calculate the contour lines shown in Figure \ref{fig:overlay}. We find that larger cores correspond to larger values of $\sqrt{\tilde{D}}/J_r$. More importantly, our estimate is that $\sqrt{\tilde{D}}/J_r \sim 0.2$ in the simulations with $\epsilon = 0.04$. According to \citet{2020arXiv201200737B}, this corresponds to the regime in which the evolution of radial action distributions transitions from adiabatic to impulsive behaviour. In particular, it can in fact be impulsive for a significant subset of particles, which would explain the signatures of impulsive energy injection seen in the orbital family for these configurations in our simulations (see also Figure \ref{fig:disk_sph_comp}).

\section{Discussion}\label{sec:discussion}
In this article, we have presented an effective model for SN feedback 
that can be adopted in DMO simulations of isolated halos in order to quickly investigate the impact of changing the total energy budget of SN feedback, the timescale over which the energy is injected, and the spatial concentration of the baryonic matter within the DM halo. 
In this Section we 
discuss some key elements and assumptions of our model. In particular, we first focus on some strengths and weaknesses of our chosen galaxy models and of how we determine the total SN feedback energy. We then discuss the differences in the results we obtain depending on the concentration of the baryonic galaxy and on how impulsive the energy deposition is. 
Eventually, we discuss the implications of our results for how we can use kinematic tracers to detect the imprint of impulsive SN feedback in dwarf galaxies.  

\subsection{Distribution of baryonic matter}\label{sec:barmetdist}
All of our simulations are of a DM halo of mass $M_{200} = 10^{10}h^{-1}{\rm M}_\sun$ with an initial concentration of $c_{200} = 13$. The halo is initially set in approximate dynamical equilibrium, having a NFW density profile for radii smaller than $r_{200}$, and an exponential cutoff at larger radii. In a first step, we add an external galactic potential to the halo and wait for the halo to contract before 
applying our SN feedback model. The parameters defining all of the six external potentials used in our work are listed in Table \ref{tab:simu_params} and described in Section \ref{sec:ICS}. The benchmark cases have been chosen in order to approximate 
Fornax (Plummer sphere) and the SMC (flat exponential disk). 

The values of $a$ and $M_{\rm Pl}$ in our benchmark Plummer sphere are inspired by measured properties of Fornax reported in \citet{2012AJ....144....4M}, as well as Fornax's historic gas content reported in \citet{2016MNRAS.456.3253Y}. Moreover, the size of our halo corresponds to roughly the measured size of Fornax's host halo. It should be noted that our Plummer model is spherically symmetric, and since Fornax is flattened (e.g. \citealt{2018MNRAS.474.1398G}), our model does not give a true representation of Fornax's gravitational potential. However, the contraction of the initial DM halo due to 
the baryonic mass estimated for Fornax is
negligible in all three configurations we explore. Critically for our purposes, a spherically symmetric potential allows for a direct investigation of whether our implementation of SN feedback is impulsive or adiabatic by following the evolution of the phase space distribution of an orbital family as presented in \citet{2019MNRAS.485.1008B}. 

Our benchmark disk model has been chosen in order to approximate the SMC following 
\citet{2012MNRAS.421.3488H}, but scaling down the mass of the disk by a factor of 2, since our host halo is smaller in mass than the assumed SMC's host halo by exactly that factor. We note that due to the same reason the scale length of our benchmark exponential disk potential should in principle be reduced as well. However, it is unclear 
how to make this correction since the presence of a prominent gaseous component (far more extended than the stellar disk) in a SMC-like galaxy complicates matters.
In order to not over-estimate the gravitational effect of the external disk, we chose to fix its scale length to the value reported in \citet{2012MNRAS.421.3488H}. Departures from this choice are nevertheless considered when we discuss the impact of the scale length on
how effective SN feedback is at forming a core. 
We note that 
since our SMC-like system is simply a scaled down version of the SMC and its host halo reported in \citet{2012MNRAS.421.3488H}, its stellar-to-halo mass ratio is somewhat inconsistent with observations \citep[e.g.][]{2010ApJ...710..903M}. 
We thus emphasize that our disk-galaxy models should be taken as a case study of how the impact of SN feedback changes if the modelled dwarf galaxy is heavier (and axisymmetric instead of spherically symmetric).  

\subsection{SN feedback energy deposition}\label{sec:en_snfeedback}
The nominal SN feedback energy injected into the DM halo from each mock galaxy is calculated from Equation \ref{eq:penarrubia}, taken from \citet{Penarrubia:2012bb}. This total energy budget depends on the choice of the initial mass function and on the effective energy coupling of SN feedback to DM ($\epsilon_{\rm DM}$). The latter remains a subject of debate 
with a broad range of values between 0 and 1 effectively used across diverse SN feedback implementations. In this work, we do not model $\epsilon_{\rm DM}$ directly, but rather measure it at the end of our simulations. To determine the mass of individual ``supernovae'', we explore the range of values between 0.01 and 0.4 for the nominal coupling of the SN feedback energy to the ISM, i.e., the values that \citet{Penarrubia:2012bb} considered plausible (see Section \ref{sec:reseng} for a discussion of our energy injection scheme and associated caveats). 
We remark that the available energy for SN feedback depends linearly on the stellar mass in the galaxy, whereas the energy required for cusp-core transformation depends on the square of the DM halo mass. Thus, it is not surprising that \citet{DiCintio2014}, \citet{Tollet2016}, \citet{Chan2015}, \citet{2017MNRAS.471.3547F} and \citet{2020MNRAS.497.2393L} find that the inner slope of DM halos in hydrodynamic simulations is a function of the stellar-to-halo mass ratio. 
In addition to Section \ref{sec:reseng}, we here make a few further remarks about 
how we implement the energy injection through SN feedback in our effective model: 
\begin{itemize}
    \item Contrary to how SN feedback occurs in hydrodynamical simulations, the ``superbubbles'' in our model are stationary over $600\,h^{-1}$Myr. This is larger than the typical orbital times at the relevant radii and may introduce artificial asymmetries into the halo. This effect can be reduced by increasing $N_{\rm SNF}$, and thus creating, on average, a more isotropic distribution of ``supernovae''.
    \item Negative local ``gas'' densities can occur in the ``superbubbles''. This is not an issue within our model unless the associated acceleration becomes large enough to disrupt the halo. In fact, a large density contrast is desirable, as it leads to a better agreement between the nominal and actual injected energy (see Section \ref{sec:reseng}).
    \item The two points above represent opposing requirements for the parameter $N_{\rm SNF}$. Having a better handle on the injected energy requires $N_{\rm SNF}$ to be small, while an increased symmetry is obtained for larger $N_{\rm SNF}$. We have checked for a few benchmark simulations that varying $N_{\rm SNF}$ by a factor of 2 up or down does not significantly affect our results.  
    \item Since our idealized simulations are initialized from halo properties today, we are not modeling the cosmological history of halo assembly. Keeping this caveat in mind, our analysis is more appropriate for dwarf galaxies with a fairly recent dominant star formation activity, some of which have been associated to have cored DM density profiles \citep{2019MNRAS.484.1401R}.
    \item The very large cores that form for $\epsilon = 0.4$ in the simulations with an external Plummer potential and short injection time (red circles and left pointing triangles in Figure \ref{fig:core_size_results}) are fairly unphysical with SN feedback 
    disrupting the halo's density profile out to radii of almost $10\,{\rm kpc}$. 
    Simulations with this combination of parameters have nevertheless provided valuable insight into the nature of the SN-driven mechanism of cusp-core transformation. 
    \item Finally, we briefly note that the parameters $f_\star$ and $\epsilon$ are degenerate in our model. We include $f_\star$ as a parameter to facilitate a comparison of our results with observations and simulation results in the literature and to relate the feedback energy to the total stellar mass.  
\end{itemize}

\subsection{Impact of the concentration of baryons on SN feedback}
We investigate the impact of how concentrated baryonic matter is within the halo in the SN-driven mechanism of cusp-core transformation by
varying the scale length $a$ ($H$) of the external Plummer (exponential disk) potential. The mass distribution corresponding to those potentials directly determines the spatial distribution of explosion centers in our SN feedback model (see Section \ref{sec:sndist}). 

In the case of the disk potential, our modelling has a couple of caveats, which could lead to underestimating the SN feedback impact. 
The first one is that in order to analytically calculate the potential, we assume the galaxy to be an infinitely flat exponential disk (see Section \ref{sec:extpot}).
However, we distribute the locations of individual ``SN centers'' in a more realistic way; they are not placed exactly in the disk plane but follow a $\cosh^{-2}$ distribution in the vertical direction. 
This makes our implementation of SN feedback slightly inconsistent with the calculation of the external potential, but we accept this inaccuracy in the interest of distributing the ``SN locations'' more realistically. The second reason we may underestimate the impact of SN feedback in the disk case is that we do not differentiate between gas and stars in the baryonic disk. 
Modelling a two-component disk potential with a more extended gaseous disk would thus reduce the
contraction of the DM halo (see Figure \ref{fig:ICgeneration}), while keeping the amount of injected SN energy, as well as its spatial distribution, the same. These effects should lead to an increased impact of SN feedback. However, as we mentioned above, the mass of our disk potential lies above the \citet{2010ApJ...710..903M} stellar-to-halo mass relation and thus the calculated energy available for SN feedback is rather large. This over-estimate of the SN feedback energy can potentially cancel some of the suppression effects outlined above. 

Irrespective of how well our results can be compared to observations of real dwarf galaxies, a clear trend emerges when changing the scales of the external disk or Plummer potential. In case of a nominal 
energy coupling $\epsilon = 0.05$, this trend becomes particularly obvious in Figure \ref{fig:core_size_results}. The more concentrated the baryonic distribution is, 
the larger the reduction in central DM density, provided the injection time is shorter than a typical dynamical time. 
This result is in general agreement with the findings of \citet{2019MNRAS.488.2387B} and \citet{2020MNRAS.497.2393L}. Namely, these works find that increasing the star formation threshold in hydrodynamical simulations leads to denser and more concentrated gas in the halo center, and subsequently to a more concentrated energy injection into the halo, making SN feedback more efficient at forming DM cores. 
This does not mean that an inversion of this trend, as reported in \citet{2019MNRAS.488.2387B}, can be disproved by our model. In our model, ``gas'' is removed from the central halo instantaneously. In hydrodynamic simulations, gas can only be removed from the center of the galaxy if the energy injected through SN feedback is sufficiently large to overcome the gravitational pull of the combined potential of the DM and the baryonic matter in the center. In the simulations of \citet{2019MNRAS.488.2387B}, this condition may not be fulfilled for very large star formation thresholds, due to the concentrated accumulation of baryonic mass in the center of the galaxy. This would explain the ``sweet spot'' range of star formation thresholds for core formation reported by the authors. While we cannot conclude from 
our results whether cores can form in very dense systems or not, we can make a different statement. If cores are formed in these systems, then DM cusps are unlikely to be restored due to the gravitational pull of gas that re-accumulates in the center of the galaxy. In other words, the DM cores that are formed in our simulations are stable -- even in the presence of a very centrally concentrated ``baryonic'' potential. 

\subsection{Impulsiveness of SN feedback}
There is an overall consensus in the literature that star formation needs to be ``bursty'' for SN feedback to be efficient at forming a DM core. In this article, we find that the most important criterion is the amount of energy that is deposited into the system by SN feedback. Nonetheless, we find that the energy injection timescale plays an important role as well, either in determining the core significance (for high energies, $\epsilon \sim 0.4$), or whether or not a core is formed at all (for low energies, $\epsilon \sim 0.05$, see Figure \ref{fig:core_size_results}). We find virtually no difference in our results (core significance and the presence of signatures of impulsive SN feedback in the final phase space distribution of an orbital family of tracers) between simulations with injection times that are $\sim 1\%$ of the dynamical time and injection times that are $\sim 10\%$ of the dynamical time. 
The situation changes significantly once the injection time becomes comparable to the dynamical time.
For small and medium energy couplings, DM cores do not form for such long injection times and no signatures of impulsive SN feedback can be detected in the phase space distribution of tracers.
The case of a large nominal energy coupling ($\epsilon=0.4$) is different, both DM cores and their signatures in the phase space of orbits are present, albeit reduced in significance.
As explained in Section \ref{sec:core_imp}, the reason for this is that if $\epsilon=0.4$, the change in the gravitational potential is substantial enough to be perceived as impulsive by particles in the halo's center, even if the SN feedback energy is injected on a timescale comparable to the dynamical time. 

Overall, we find that signatures of impulsive SN feedback can always be detected in the phase space distribution of an orbital family of tracers if the DM halo has formed a core. The reverse statement, however, is not true. If the energy coupling is rather low ($\epsilon = 0.01$) cores do not form (see Figure \ref{fig:core_size_results}), yet we can see some signatures of an impulsively changing gravitational potential imprinted in the phase space of the orbital family, provided the energy is injected on timescales shorter than the dynamical time. In general, we find that while these signatures are present, the net radial expansion and diffusion of orbits is less significant for SN feedback that is not energetic enough to form a core. Still, this raises the question of whether and how such kinematic signatures can be used to differentiate between adiabatic and impulsive core formation, as suggested in \citet{2019MNRAS.485.1008B}. One possibility is to look at differences in age and metallicity gradients of stars between a system that undergoes adiabatic core formation and one that undergoes impulsive core formation. We are currently investigating this issue (Burger et.al., in prep.) by means of a suite of hydrodynamical simulations of a single isolated halo, including different star formation thresholds 
using the stellar feedback model {\scriptsize SMUGGLE} \citep{2019MNRAS.489.4233M} 
incorporated into {\scriptsize AREPO} \citep{Springel:2009aa}.

\section{Summary}\label{sec:conclusion}

We have presented a new effective model of SN-driven cusp-core transformation that can be included in $N$-body simulations of isolated halos.
Our model consists of two main components, an external potential that approximates the distribution of baryons in a dwarf galaxy, and a scheme to inject energy into the DM particle distribution in a manner that is approximately consistent with the stellar distribution modelled by the external potential. In a series of simulations, we have tested how the effect of SN feedback depends on the baryonic concentration, the amount of injected SN feedback energy, and the timescale on which this energy is injected into the halo. We have used simulations of a dwarf-size halo to examine the cases of a Plummer potential (to mimic a Fornax-like system)
and a disk potential (to mimic a SMC-like system). 
We find that the most important factor determining whether SN feedback can form cores in dwarf galaxies is whether or not enough energy is available to transform the halo's density profile. If the available energy is close to maximal, cores form even if the SN injection time is longer than the dynamical timescale in the halo center and/or baryons are concentrated or not. If less energy is available, whether or not cores form depends on how fast the energy is injected and on how concentrated the baryonic matter is within the halo. For minimal values of the energy, 
cores cannot form. 
For a fixed amount of feedback energy, larger cores form for faster injection times and more concentrated galaxies. Cores formed in very concentrated galaxies are stable -- adiabatic contraction due to the centrally concentrated baryonic potential does not restore the cusp, even if no further supernovae occur. 
Analyzing the phase space distribution of tracer particles, we find clear signatures of impulsive SN feedback in all simulations in which the DM halo develops a core. However, we also find these signatures in a few simulations in which the halo's density profile remains cuspy as the amount of SN feedback energy is insufficient to trigger core formation. The longevity and the appearance of those signatures are closely linked to the spatial symmetry of the halo and the external potential. 

\acknowledgments

JB and JZ acknowledge support by a Grant of Excellence from the
Icelandic Research Fund (grant number 173929). The simulations in
this paper were carried out on the Garpur supercomputer,
a joint project between the University of Iceland and University of Reykjav\'ik with funding from the Icelandic Research Fund.

\bibliography{manuscript}{}
\bibliographystyle{aasjournal}



\end{document}